\documentclass[amsmath,reprint,prb,longbibliography,superscriptaddress,bibnotes]{revtex4-2}
\usepackage{amsthm,amssymb,amsfonts,graphicx,verbatim, xcolor,bm} 
\usepackage{hyperref} 
\usepackage[utf8]{inputenc} 
\usepackage{siunitx} 
\graphicspath{{figures/}}

\newcommand{\ket}[1]{\lvert #1 \rangle}

\begin{document}

\title{Composite Fermion Theory of Fractional Chern Insulators in Rhombohedral Graphene}
\author{Peize Ding}
\affiliation{Department of Physics, Columbia University, New York, NY 10027}
\author{Andrew J. Millis}
\affiliation{Department of Physics, Columbia University, New York, NY 10027}
\affiliation{Center for Computational Quantum Physics, Flatiron Institute, New York, NY 10010}
\author{Yongxin Zeng}
\email{yongxin.zeng@yale.edu}
\affiliation{Department of Physics, Columbia University, New York, NY 10027}
\affiliation{Department of Physics, Yale University, New Haven, CT 06520}

\begin{abstract}
The experimental observation of the fractional quantum anomalous Hall effect in rhombohedral graphene multilayers has raised questions about the nature of the fractional Chern insulator (FCI) states and the role of the moir\'e potential. In this Letter we develop a composite fermion (CF) description and discuss how moir\'e modulation and Coulomb interactions stabilize the FCI ground states. 
Under a flux-attachment transformation and mean-field approximation, the CF spectrum forms Landau levels that are further split by the moir\'e potential. At moir\'e filling fractions in the Jain sequence, an integer number of CF bands are occupied, resulting in an incompressible ground state with fractionally quantized Hall response.
Remarkably, when the attached flux opposes the Berry curvature of the conduction band of rhombohedral graphene, the CF Landau levels are widely spaced in energy and delocalized across layers, leading to a reduced density of states that helps protect the CF mean-field state and amplifying effects of the moir\'e potential acting on the layer away from the doped electrons. Interactions between CFs further enlarge the moir\'e-induced band gap and stabilize the FCI ground state.
Our work sheds light on the nature of the FCI states in rhombohedral graphene and provides a new theoretical framework for further studies.
\end{abstract}

\maketitle

{\it Introduction.---}The fractional quantum Hall (FQH) effect \cite{stormer1999fractional} is one of the most spectacular phenomena in condensed matter physics arising from the interplay of quantum geometry, topology, and strong electron-electron interactions. Fractional Chern insulators (FCIs), which display many of the features of FQH states but exist in energetically isolated topological flat bands in the absence of external magnetic field, were theoretically proposed \cite{regnault2011fractional, sheng2011fractional, neupert2011fractional, tang2011high, sun2011nearly} over a decade ago and were recently observed experimentally first in twisted MoTe$_2$ bilayers \cite{cai2023signatures, park2023observation, zeng2023thermodynamic, xu2023observation} and subsequently in electron-doped rhombohedral $L$-layer graphene (R$L$G for $L=4,5,6$) encapsulated with hexagonal boron nitride (hBN) \cite{lu2024fractional, lu2025extended, xie2025tunable, waters2025chern, aronson2025displacement, choi2025superconductivity, butler2026fci}. The latter observations were unexpected because on the level of standard band theory the moir\'e potential produced by the hBN angle-aligned with the graphene layers does not produce a modulation potential strong enough to create an {\it isolated} topological flat band. A further surprise is that the FCIs observed in R$L$G appear more robust when the moir\'e potential acts on the layer farthest away from the doped electrons, when it is expected to produce the weakest modulation of the electronic band structure; when the moir\'e potential acts on the layer with the highest electron density, other insulating phases are more often observed \cite{aronson2025displacement}. To reconcile this, several groups have proposed \cite{dong2024anomalous, dong2024theory, zhou2024fractional, guo2024fractional} that the doped electrons undergo spontaneous crystallization, and the resulting ``anomalous Hall crystal" \cite{zeng2024sublattice, tan2024parent, soejima2024anomalous, dong2024stability, tan2025variational, zhou2025new, kwan2025moire, bernevig2025berry, desrochers2026energetics, maymann2026skyrmion} is the parent state of the FCIs. However, more recent numerical studies \cite{yu2025moire, li2025multi, regnault2026moire} find that these FCI states are not robust against band mixing effects. Moreover, no integer or fractional quantum anomalous Hall states have been observed in R$L$G in the absence of moir\'e \cite{han2025signatures, seo2025family, han2026evidence}, suggesting that the translational symmetry breaking required to define the FCI states is imposed, not spontaneous. The nature of the FCI states and the role of moir\'e have remained under debate.

Theoretical studies of FCIs have traditionally relied on analytic construction of Laughlin-type many-body wave functions in ``ideal band" models \cite{roy2014band, jackson2015geometric, claassen2015position, wang2021exact, ledwith2023vortexability} or numerical exact-diagonalization studies of finite-size systems \cite{regnault2011fractional, sheng2011fractional, neupert2011fractional}. The starting point of both approaches is a projection to a single, narrow band which is a valid approximation for systems like twisted MoTe$_2$ but is inappropriate for R$L$G. In this paper, we consider an alternative approach: the composite fermion (CF) theory \cite{jain1989composite, jain2007composite, jain2015composite, lopez1991fractional, halperin1993theory}. In its original form developed for FQH systems, new CF quasiparticles are defined by attaching an even number of flux quanta to each electron; in this quasiparticle representation the external magnetic field is effectively reduced and the FQH effect of electrons is transformed into the integer quantum Hall effect of CFs. Despite the insights it has provided in the FQH context, the CF approach to zero-field FCIs has been much less explored until very recently \cite{sohal2018cf, lu2024CF, gassner2025flux, hu2026cf, jin2026cf}. 
We adopt this approach here because the flux-attachment transformation acts on the electron field operator and does not require an isolated narrow band, providing a natural approach for FCIs in R$L$G.

\begin{figure}
    \centering
    \includegraphics[width=\linewidth]{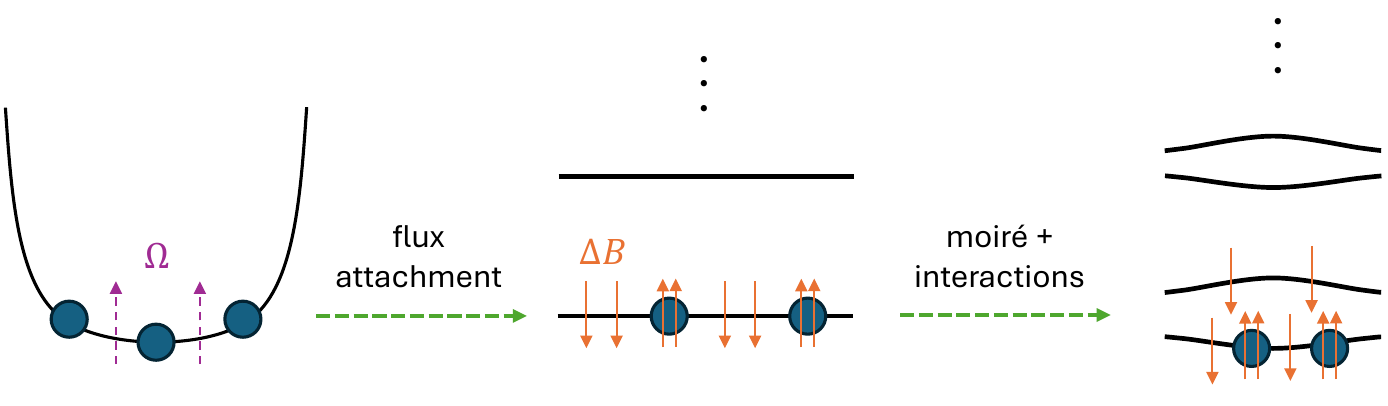}
    \caption{Schematic illustration of the composite fermion (CF) theory of fractional Chern insulators (FCIs) in rhombohedral graphene. Left panel: schematic of band structure of R$L$G in the absence of moir\'e, with orientation of Berry curvature $\Omega$ indicated. Middle panel: flux attachment maps electrons into CF quasiparticles in a magnetic field such that the lowest CF Landau level (LL) is half filled. When the Chern-Simons gauge field ($\Delta B$) is chosen to be opposite to the Berry curvature $\Omega$ of the parent band, the LLs are widely spaced in energy, stabilizing the CF mean-field state. Right panel: a periodic moir\'e potential splits each LL into Hofstadter subbands. For moir\'e periodicities such that the number of electrons per moir\'e unit cell is in the Jain sequence $\nu = n/(2n\pm 1)$, each LL splits into $2n$ subbands and the lowest $n$ are occupied (shown for $n=1, \nu=1/3$). Interactions between CFs further increase the energy gap and additionally stabilize the FCI ground state. }
    \label{fig:cartoon}
\end{figure}

In this Letter we perform a flux-attachment transformation and, at the mean-field level, examine the effects of the moir\'e potential and Coulomb interactions on the CF energy spectrum. Our approach is similar to that used by Ref.~\cite{lu2024CF} for a lattice model of twisted MoTe$_2$, but here applied to the continuum model of R$L$G and extended to include interaction effects. Fig.~\ref{fig:cartoon} summarizes our core physical picture: the combination of flux attachment and translation symmetry breaking (via moir\'e modulation and interactions) leads to an incompressible CF state. We find that when the attached magnetic field is opposite to the Berry curvature of the electronic parent band, the density of states is reduced and the effect of the moir\'e potential away from the doped electrons is greatly enhanced compared to its negligible effect on the electronic band structure. Interactions between CFs further increase the energy gap and stabilize the FCI ground state. 

{\it Model.---}We consider R$L$G in a displacement field and describe it using a $2L$-band continuum model in layer-sublattice space:
\begin{align} \label{eq:H_RLG}
    H_{{\rm R}L{\rm G}} = &\sum_{ll'\sigma\sigma'\bm k} [h(\bm k)]_{l'\sigma',l\sigma} c_{l'\sigma'\bm k}^{\dagger} c_{l\sigma\bm k} \nonumber\\
    &+ \sum_{l\sigma\bm k} (\frac{L+1}{2}-l)u_D c_{l\sigma\bm k}^{\dagger} c_{l\sigma\bm k},
\end{align}
where $l=1,\dots,L$ is the layer index and $\sigma=A,B$ is the sublattice index, $c^{\dagger}$ and $c$ are electron creation and annihilation operators, and $u_D$ is the interlayer potential difference induced by the displacement field. The $2L\times 2L$ matrix $h(\bm k)$ encodes intralayer and interlayer hoppings in graphene and its precise form is provided in the Supplemental Material (SM) \cite{SM}. $H_{{\rm R}L{\rm G}}$ describes spin-polarized electrons within a single valley of R$L$G; the opposite spin and valley are neglected by assuming spontaneous spin and valley polarization as is widely observed in experimentally relevant parameter regimes. For concreteness, we select the valley in which the Berry curvature near the conduction band bottom is positive. 

The moir\'e superlattice formed at the interface between an outer graphene layer ($l_0=1$ or $L$) and angle-aligned hBN layers is modeled phenomenologically as a periodic modulation potential acting on layer $l_0$:
\begin{equation}
    H_{\textrm{moir\'e}} = \sum_{\sigma\bm{Gk}} V_{\bm G}^m c_{l_0 \sigma,\bm k+\bm G}^{\dagger} c_{l_0 \sigma\bm k}.
\end{equation}
Here $\bm G$ represents reciprocal lattice vectors of the triangular moir\'e lattice and $V_{\bm G}^m$ are the Fourier coefficients of the modulation potential. In our explicit calculations we focus on pentalayer graphene (R5G) and use the first-harmonic approximation so that $V_{\bm G}$ is nonzero only for $|\bm G| = G_1 \equiv 4\pi/\sqrt{3}a_m$ ($a_m$: moir\'e lattice constant) and the moir\'e potential is parameterized by a single complex number $V_m \equiv V_{(G_1,0)}^m = |V_m| e^{i\phi}$. Microscopic models of the graphene/hBN moir\'e also include off-diagonal terms \cite{park2023topological, herzog2024moire} and sometimes a modulation on all layers \cite{regnault2026moire}, but we omit these as they do not affect the qualitative physics we discuss below.

\begin{figure}[b]
    \centering
    \includegraphics[width=1.0\linewidth]{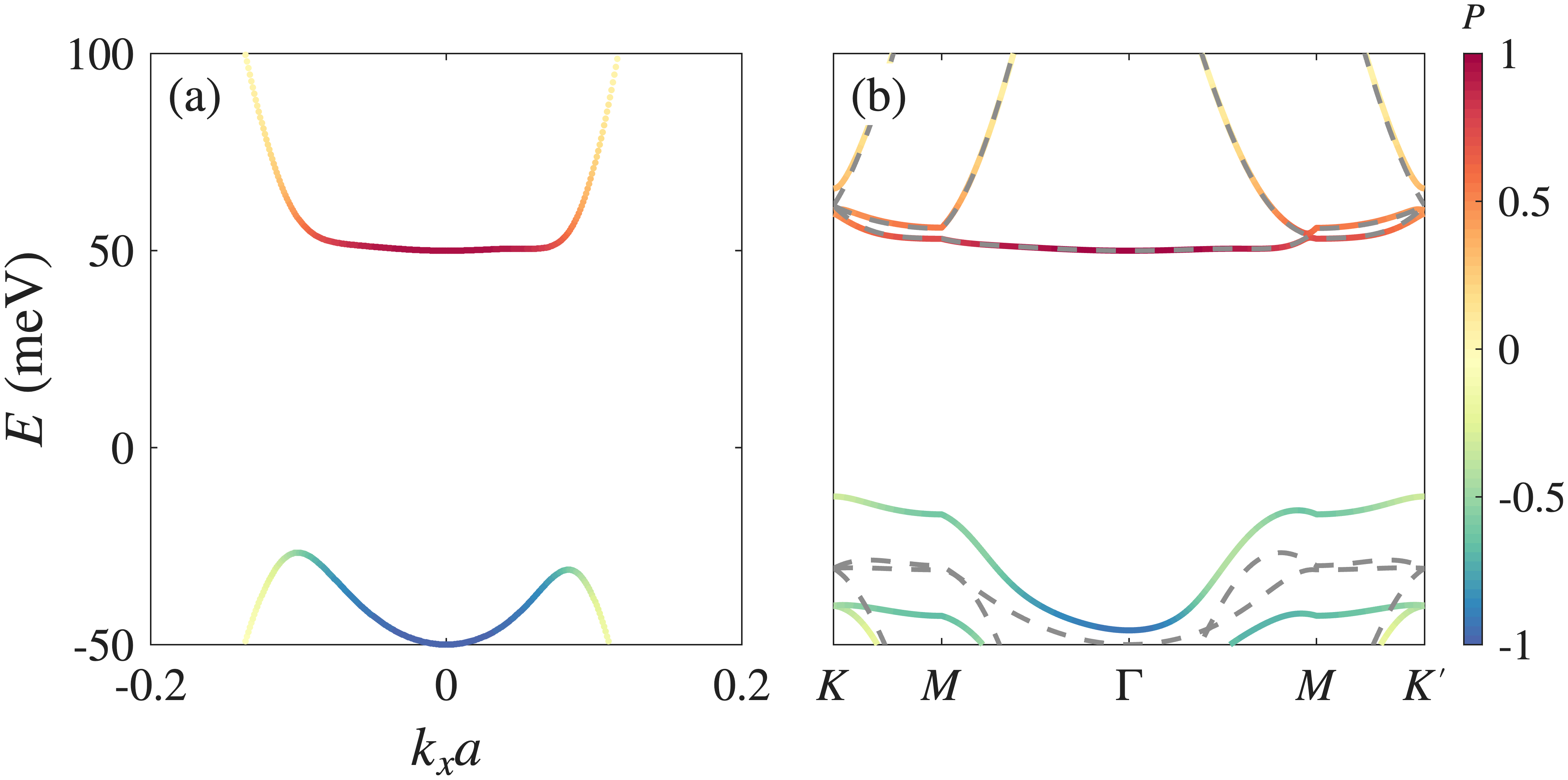}
    \caption{Electronic band structure of rhombohedral pentalayer graphene (R5G). (a) Continuum band structure in the absence of moir\'e. (b) Moir\'e band structure along $K$--$M$--$\Gamma$--$M$--$K'$ for $V_m=\SI{20}{meV}$ applied to layer $l_0=5$. The gray dashed lines show the $V_m=0$ bands folded into the moir\'e Brillouin zone. The color scale indicates the layer polarization $P$ [Eq.~\eqref{eq:P}], defined such that $P=+1$ ($-1$) when an electronic state is fully polarized to layer $l=1$ ($l=5$).}
    \label{fig:band}
\end{figure}

Fig.~\ref{fig:band}(a) shows the electronic band structure of R5G without moir\'e calculated from Eq.~\eqref{eq:H_RLG} at $u_D = \SI{25}{meV}$, while Fig.~\ref{fig:band}(b) shows how moir\'e modulation on layer $l_0=5$ reconstructs the low-energy states. The color of the bands reflects their layer polarization defined as
\begin{equation} \label{eq:P}
    P \equiv \sum_{l=1}^L \frac{L+1-2l}{L-1} |\psi_l|^2,
\end{equation}
so that $P=+1$ when the electron Bloch state is fully polarized to layer $l=1$ and $P=-1$ when it is polarized to $l=L$. Here $|\psi_l|^2$ is the probability density of the electron Bloch state in layer $l$.
The displacement field opens a large gap in the band structure, and the conduction band above the gap is nearly dispersionless over the momentum range $ka \lesssim 0.1$ ($a$: graphene lattice constant). The electronic states at the flat band bottom are strongly polarized to layer $l=1$. Consequently, when the moir\'e potential acts on the remote layer $l_0=5$, the conduction band dispersion remains largely unperturbed. Throughout this work, we fix the moir\'e parameters to $a_m = \SI{11.5}{nm}$, $|V_m| = \SI{20}{meV}$, and $\phi=0$. Here the value of $a_m$ is chosen to match experimental conditions, whereas $|V_m|$ and $\phi$ serve as phenomenological parameters whose precise values do not alter our qualitative conclusions.

Electron-electron interactions in R$L$G are described by the Hamiltonian
\begin{equation} \label{eq:H_int}
    H_{\rm int} = \frac{1}{2A} \sum_{ll'\sigma\sigma'} \sum_{\bm{kk'q}} V_q c_{l\sigma,\bm{k+q}}^{\dagger} c_{l'\sigma',\bm{k'-q}}^{\dagger} c_{l'\sigma'\bm{k'}} c_{l\sigma\bm{k}},
\end{equation}
where $A$ is the area of the two-dimensional system and $V_q = 2\pi e^2/\epsilon q$ is the Coulomb potential. In our numerical calculations we set the dielectric constant to $\epsilon=5$, and it follows that the interaction energy scale is $e^2/\epsilon a_m = \SI{25}{meV}$.

{\it Flux attachment.---}In the absence of interactions, a fractionally filled Landau level (LL) has an extensive ground-state degeneracy, which is lifted by interactions leading to FQH states. Similarly, the very weak dispersion of the lowest conduction band in RLG leads to an extremely large density of state, implying very large interaction effects and a reconstructed many-body ground state. A natural basis for describing these fractionalized states is provided by CFs, constructed via a singular gauge transformation on the electron operators
\begin{equation} \label{eq:flux_attachment}
    \psi_{\rm cf}^{\dagger}(\bm r) \equiv \psi_e^{\dagger}(\bm r) \exp\left[-2is\int d^2 \bm r' \arg(\bm r-\bm r') \rho(\bm r')\right]
\end{equation}
that binds an even integer number ($2s$) of magnetic flux quanta to each electron. Here $\rho(\bm r) \equiv \psi_e^{\dagger}(\bm r) \psi_e(\bm r) = \psi_{\rm cf}^{\dagger}(\bm r) \psi_{\rm cf}(\bm r)$ is the local particle density operator. In R$L$G, we apply this transformation exclusively to the active electrons in the conduction band, while the filled valence bands are treated as an inert background that is subtracted from the definition of $\rho(\bm r)$. The flux-attachment transformation \eqref{eq:flux_attachment} preserves the fermionic exchange statistics, but the CFs are coupled to a dynamical gauge field proportional to the local particle density (see SM \cite{SM} for a detailed derivation). At the mean-field level, the dynamical gauge field is replaced by a static and spatially uniform magnetic field proportional to the average carrier density: $\Delta B = (4\pi s \hbar/e) \langle\rho\rangle$. 

\begin{figure}
    \centering
    \includegraphics[width=1\linewidth]{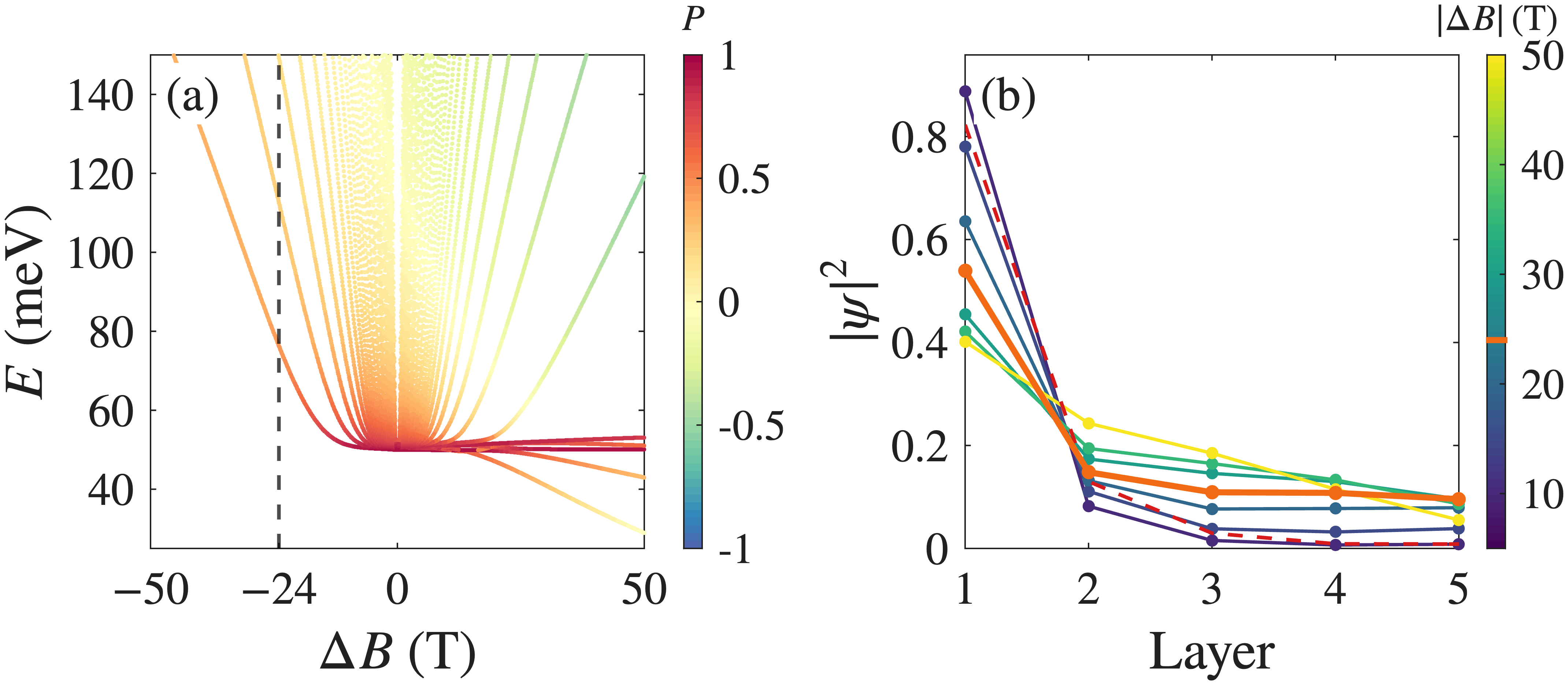}
    \caption{(a) Landau level (LL) spectrum of CFs in R5G without moir\'e, plotted as a function of magnetic field $\Delta B$. The sign convention is such that $\Delta B<0$ when the magnetic field is opposite to the Berry curvature at the conduction band bottom. The black dashed line marks $\Delta B=\SI{-24}{T}$, corresponding to the electron density at moir\'e filling $\nu=1/3$ which we focus on below. (b) Layer distribution of the lowest conduction-band LL at various magnetic field strengths $\Delta B<0$ (indicated by color), with the $|\Delta B|=\SI{24}{T}$ curve highlighted in orange. The red dashed line denotes the zero-field layer distribution of the electronic Bloch states averaged over the first moir\'e Brillouin zone. 
    }
    \label{fig:LL}
\end{figure}

{\it CF spectrum.---} In conventional LL systems, $\Delta B$ effectively reduces the external magnetic field, and the FQH effect occurs when the electron LL filling factor is in the Jain sequence $\nu = n/(2sn \pm 1)$ so that CFs occupy an integer number ($n$) of effective LLs in the reduced field. In the absence of external magnetic field, the flux-attachment transformation \eqref{eq:flux_attachment} {\it creates} an effective magnetic field. In the absence of moir\'e, the gauge field $\Delta B$ discretizes the continuum electronic band structure of R5G [Fig.~\ref{fig:band}(a)] into a set of CF LLs, plotted in Fig.~\ref{fig:LL}(a) as a function of $\Delta B$. By construction the lowest CF Landau level (LLL) is filled by $1/2s$ since we have attached $2s$ flux quanta to each electron. The level structure changes dramatically under reversal of the field direction because of the broken time-reversal symmetry of the single-valley system. For $\Delta B<0$, i.e., when the effective magnetic field is opposite to the Berry curvature of the parent conduction band, the LLs above the band gap are widely spaced in energy, while for $\Delta B>0$ the lowest five LLs above the gap are nearly degenerate. This field-induced modification of the density of states is a consequence of the momentum-space chirality of rhombohedral graphene and is consistent with the semiclassical theory of the Berry-phase correction to the density of states \cite{xiao2005berry, xiao2010berry}. Since the goal of our flux-attachment transformation is to reduce the large density of states at the flat electronic band bottom, in the following we fix the direction of the attached flux to $\Delta B<0$. This choice also leads to the correct sign of $\sigma_{xy}$ expected for a parent band with positive Berry curvature.

In the presence of a moir\'e potential, flux attachment produces Hofstadter bands. At moir\'e filling factor $\nu$, the induced flux threading each moir\'e unit cell equals $2s\nu$ times the magnetic flux quantum. For filling fractions within the Jain sequence, $2s\nu = 2sn/(2sn \pm 1)$. Because $2sn$ and $2sn\pm 1$ are coprime integers, the emergent magnetic translation symmetry dictates that each moir\'e band splits into $2sn\pm 1$ magnetic subbands, and the CF filling factor is an integer: $\nu\cdot (2sn\pm 1) = n$. If the $n$th and $(n+1)$th bands are separated by a gap, filling the lowest $n$ CF bands leads to an incompressible liquid state with electronic Hall conductance $\sigma_{xy} = -[C/(2sC+1)](e^2/h)$ where $C$ is the total Chern number of the occupied CF bands.

In this Letter we focus on the principal Jain sequence ($s=1$) which contains all the filling fractions $\nu=n/(2n\pm1)$ at which FCIs have been observed in R$L$G. We present results mainly for the simplest fraction $\nu=1/3$ ($n=1$) and leave other fractions for the SM \cite{SM}. At $\nu=1/3$ with moir\'e periodicity $a_m = \SI{11.5}{nm}$, the attached magnetic field is $\Delta B = -\SI{24}{T}$. As shown in Fig.~\ref{fig:LL}(a), such a strong $\Delta B$ produces widely spaced LLs so that the moir\'e potential may be treated as a perturbation.

Another important feature of the CF LLs is the reduced layer polarization of the LLL as shown in Fig.~\ref{fig:LL}(b). As the magnetic field $-\Delta B>0$ increases, the LLL wave function, initially localized in layer $l=1$, quickly spreads into the other layers. More explicitly, the LLL state is a superposition of LLs in different layers and sublattices. When trigonal warping terms in the Hamiltonian \eqref{eq:H_RLG} are neglected, the LLL wave function takes the approximate form
\begin{align} \label{eq:psi_LLL}
    \ket{\psi_{\rm LLL}} \approx &\psi_{1A} \ket{0}_{1A} + \psi_{1B} \ket{1}_{1B} + \psi_{2A} \ket{1}_{2A} + \psi_{2B} \ket{2}_{2B} \nonumber\\
    &+ \dots + \psi_{5A} \ket{4}_{5A} + \psi_{5B} \ket{5}_{5B},
\end{align}
where $\ket{n}_{l\sigma}$ denotes the $n$th LL orbital in layer $l$ and sublattice $\sigma$, and $\psi_{l\sigma}$ are the corresponding coefficients. The delocalization of the LLL wave function and the high LL indices in remote layers suggest that a moir\'e potential applied to a remote layer can have an important influence on CFs, despite having a virtually negligible impact on the low-energy electronic states.

Physically, the reduced layer polarization of the CF LLL indicates that the FCI states constructed below are built from not only the electronic Bloch states at the band bottom, but also higher-energy states that are spread more evenly across layers. The layer polarization of the CF LLL is in fact close to that of the electronic Bloch states at \SI{10}{meV} above the conduction band bottom. This sets the scale of kinetic energy the system must pay to form the CF state and thus provides an order-of-magnitude estimate of the interaction strength required to stabilize the FCI ground state.

\begin{figure}
    \centering
    \includegraphics[width=\linewidth]{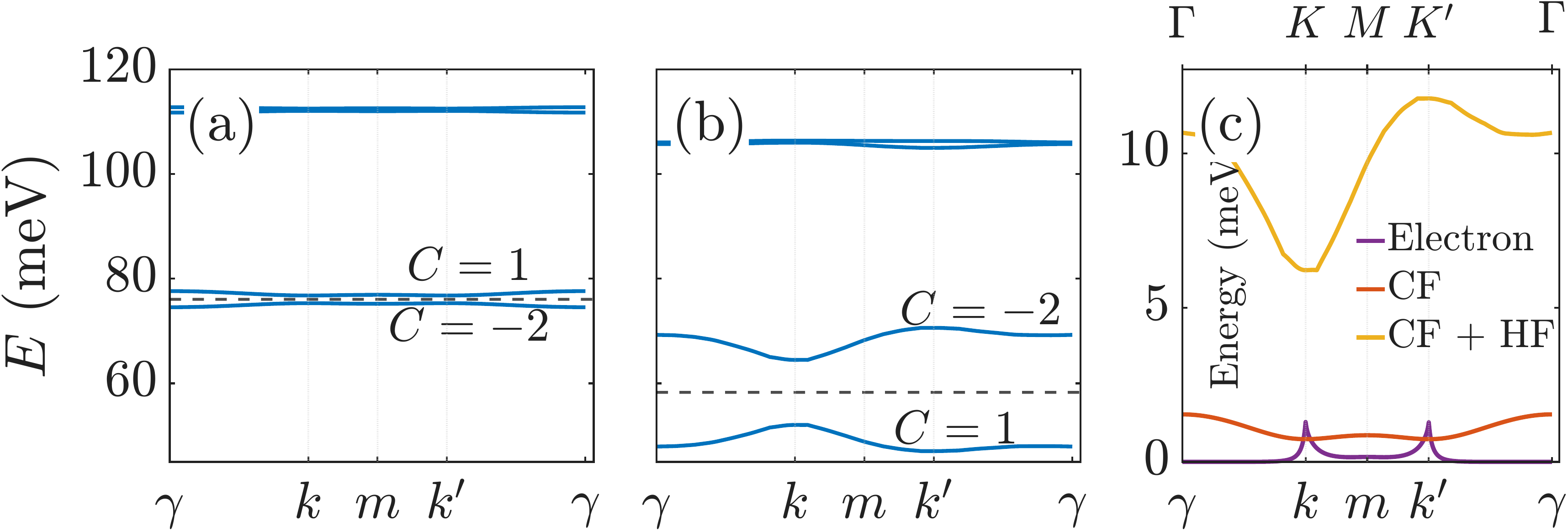}
    \caption{CF spectra and energy reconstruction at $\nu=1/3$, with $V_m=\SI{20}{meV}$ applied to the remote layer $l_0=5$ and dielectric constant $\epsilon=5$. (a) Noninteracting CF magnetic Bloch bands, where the two LLL subbands carry $C=-2$ and $C=1$. The dashed line marks the chemical potential. (b) Converged Hartree-Fock (HF) bands showing the interaction-driven inversion that places the $C=1$ band below the chemical potential. (c) Moir\'e and interaction effects on electrons and CFs. The purple curve shows the absolute moir\'e-induced energy shift of the lowest electron conduction miniband in the moir\'e Brillouin zone. The orange and yellow curves show half the direct gap between the two CF bands, without and with HF interactions, respectively.}
    \label{fig:CF_band}
\end{figure}

Fig.~\ref{fig:CF_band}(a) shows the CF band structure at $\nu=1/3$, where the LLL splits into two subbands. A remote moir\'e potential of magnitude $V_m=20\,$meV opens a gap of approximately $2\,$meV between them. While apparently small, this gap is much larger than the moir\'e-induced electronic band energy shift [Fig.~\ref{fig:CF_band}(c)], and it is comparable to the LLL splitting when the same moir\'e potential is applied to the majority electron layer ($l=1$) (see SM \cite{SM}).

{\it CF interactions.---}Despite its strong enhancement in the CF basis, the moir\'e-induced gap remains well below the interaction energy scale $e^2/\epsilon a_m = \SI{25}{meV}$, so CF interactions and band mixing cannot be neglected. Since the density operator $\rho(\bm r)$ is invariant under the flux-attachment transformation \eqref{eq:flux_attachment}, CFs interact via the same Coulomb potential \eqref{eq:H_int} as electrons. We treat CF interactions within the self-consistent Hartree-Fock (HF) approximation (see SM \cite{SM} for details). Fig.~\ref{fig:CF_band}(b) shows the resulting quasiparticle bands: interactions enlarge the gap between the two LLL subbands to about \SI{15}{meV}. However, unlike conventional LL systems in which every CF LL carries the same Chern number $|C|=1$, here the two CF subbands carry different Chern numbers that must add up to the Chern number $C=-1$ of the parent LLL; in the HF ground state [Fig.~\ref{fig:CF_band}(b)], $C=+1$ for the lower band and $C=-2$ for the upper band. Occupying the lower band gives an FCI ground state with Hall conductance $\sigma_{xy} = -e^2/3h$, in agreement with the experimental observation at $\nu=1/3$ \cite{butler2026fci}.

This agreement, however, is not a robust prediction of the CF mean-field theory. Already at the non-interacting level, which of the two CF subbands lies lower depends on the moir\'e parameter $\phi$ and the LL form factors; for the parameters we use in Fig.~\ref{fig:CF_band}(a), the lower CF band carries $C=-2$, so interactions invert the ordering. In the SM \cite{SM}, we extend this analysis to other filling fractions in the Jain sequence and find that the non-interacting CF bands correctly predict the measured Hall conductance $\sigma_{xy} = -\nu e^2/h$ for some fractions but fail for others. Coulomb interactions push the competition toward the experimentally observed state at $\nu=1/3$, but not at all fractions. The mean-field CF band ordering therefore cannot universally predict the correct topological response, and definitively resolving this energetic competition requires dynamical Chern-Simons gauge fluctuations that the current mean-field framework neglects. We leave a rigorous treatment of these fluctuations to future work.

The SM \cite{SM} presents the layer-resolved real-space charge density of the CF HF ground state. While the majority layer ($l=1$) carries the highest average charge density, its spatial modulation is relatively weak. Remarkably, the amplitude of charge density fluctuations in the remote layer ($l=5$) is comparable to that in layer $l=1$. This counterintuitive behavior stems from the distinct LL form factors of the different layer-sublattice components of the LLL wave function \eqref{eq:psi_LLL}: the higher LLs in layer $l=5$ have stronger tendency toward translation symmetry breaking. The large charge density fluctuations on the remote layer imply that a moir\'e potential acting on that layer strongly stabilizes the FCI ground state. 

{\it Discussion.---}The precise role of the moir\'e potential in the emergence of FCIs in R$L$G has been a subject of intense theoretical debate. Our CF framework shows that both the moir\'e potential and Coulomb interactions are indispensable. Although spontaneous translation symmetry breaking is allowed in principle, a CF crystal is unlikely to be the ground state of a half-filled CF LLL without the assistance of an external moir\'e potential. The ground states observed in the absence of moir\'e are instead electronic Fermi liquids and chiral superconductors \cite{han2025signatures, seo2025family, han2026evidence}, which translate into the composite Fermi liquid \cite{halperin1993theory} and Moore-Read states \cite{moore1991nonabelions} of CFs. By opening a sizable energy gap in the otherwise half-filled CF LLL, the remote moir\'e potential stabilizes the incompressible FCI ground state against these competing electronic phases. When the moir\'e potential acts instead on the majority electron layer, it opens a CF gap of similar magnitude but also strongly modulates the electronic bands \cite{SM}, so it may favor electronically ordered phases such as topologically trivial charge density waves. This may explain why FCIs appear more robust when the moir\'e potential acts on the remote layer.

The CF mean-field treatment does, however, introduce an artificial particle-hole symmetry between the $C=+1$ and $C=-2$ states within the CF LLL. In the electron picture, these are $1/3$ and $2/3$ FCI states at the same moir\'e filling $\nu=1/3$, which are not related by any symmetry of the electron Hamiltonian. The moir\'e potential does break this symmetry, but we do not expect it to be the decisive factor since the resulting splitting is small, and the selection it makes depends on the moir\'e parameter $\phi$ and predicts a $\sigma_{xy}$ that varies non-monotonically with electron density \cite{SM}, contrary to experiment. We expect instead that inter-LL Coulomb interactions and Chern-Simons fluctuations break this emergent symmetry and select the true ground state. In a Dirac system, the fluctuation term takes the form of a two-body interaction $H_{\rm fluc} = \int d^2 r :\bm j\cdot \delta\bm a:$ ($\bm j$: current operator; $\delta\bm a$: fluctuating gauge field) and does not include a three-body term as in conventional quadratic-band systems. Since $H_{\rm fluc}$ arises from the kinetic Hamiltonian, its matrix elements scale with the same energy scale that sets the CF LL spacing. The wide separation of the CF LLs therefore does not suppress inter-LL mixing, and multiple CF LLs must be retained.

The CF framework provides a systematic recipe for constructing the many-body ground state wave function. Computing its total energy would establish quantitatively whether this state is favored over those from previous theoretical proposals such as the anomalous Hall crystal picture. Another test is to apply the CF construction to closely related but numerically tractable models, such as superlattice-modulated bilayer graphene \cite{ghorashi2023topological, zeng2024gate} and the Berry basin model \cite{zeng2026basin}, where the CF wave functions can be compared directly against exact diagonalization results. Investigations of instabilities of the states considered here towards other states both in the proximal and remote moir\'e cases would also be of interest. We leave these numerical investigations for future work.

{\it Acknowledgments.---}We thank Bo Zou, Nicol\'as Morales-Dur\'an, Jennifer Cano, and Nicolas Regnault for helpful discussions.
P.D. was
supported by the National Science Foundation (NSF)
under CAREER Award No. DMR-2340394, the Materials Research Science and Engineering Centers (MRSEC) program through Columbia University under the
Precision-Assembled Quantum Materials (PAQM) Grant
No. DMR-2011738.
Y.Z. and A.J.M. acknowledge support from Programmable Quantum Materials, an Energy Frontiers Research Center funded by the U.S. Department of Energy (DOE), Office of Science, Basic Energy Sciences (BES), under award DE-SC0019443.
Work at Yale was supported by the Air Force Office of Scientific Research (AFOSR) under Award No. FA95502510287  and by  NSF Grant No. DMR-2410182.
The Flatiron Institute is a division of the Simons Foundation.

\bibliography{ref}

@misc{SM,
    note = "See Supplemental Material, which includes Refs.~\cite{zhou2021half, kolavr2024hofstadter, zou2025valley, murthy2003hamiltonian}, for detailed derivation of the flux-attachment mean-field formalism and supplemental results at other filling factors and in moir\'e-proximate systems."
}

@article{zeng2026basin,
  title={Integer and Fractional Chern Insulators in a ``Berry Basin" Model},
  author={Zeng, Yongxin and Millis, Andrew J},
  journal={to appear},
  year={2026}
}

@article{ghorashi2023topological,
  title = {Topological and Stacked Flat Bands in Bilayer Graphene with a Superlattice Potential},
  author = {Ghorashi, Sayed Ali Akbar and Dunbrack, Aaron and Abouelkomsan, Ahmed and Sun, Jiacheng and Du, Xu and Cano, Jennifer},
  journal = {Phys. Rev. Lett.},
  volume = {130},
  issue = {19},
  pages = {196201},
  numpages = {8},
  year = {2023},
  month = {5},
  publisher = {American Physical Society},
  doi = {10.1103/PhysRevLett.130.196201},
  url = {https://link.aps.org/doi/10.1103/PhysRevLett.130.196201}
}

@article{zeng2024gate,
  author       = {Zeng, Yongxin and Wolf, Tobias M. R. and Huang, Chunli and Wei, Nemin and Ghorashi, Sayed Ali Akbar and MacDonald, Allan H. and Cano, Jennifer},
  title        = {Gate-tunable topological phases in superlattice modulated bilayer graphene},
  journal      = {Phys. Rev. B},
  volume       = {109},
  number       = {19},
  pages        = {195406},
  year         = {2024},
  month        = {5},
  doi          = {10.1103/PhysRevB.109.195406},
}

@article{zhou2021half,
  title={Half-and quarter-metals in rhombohedral trilayer graphene},
  author={Zhou, Haoxin and Xie, Tian and Ghazaryan, Areg and Holder, Tobias and Ehrets, James R and Spanton, Eric M and Taniguchi, Takashi and Watanabe, Kenji and Berg, Erez and Serbyn, Maksym and others},
  journal={Nature},
  volume={598},
  number={7881},
  pages={429--433},
  year={2021},
  publisher={Nature Publishing Group UK London}
}

@article{kolavr2024hofstadter,
  title={Hofstadter spectrum of Chern bands in twisted transition metal dichalcogenides},
  author={Kol{\'a}{\v{r}}, Kry{\v{s}}tof and Yang, Kang and von Oppen, Felix and Mora, Christophe},
  journal={Physical Review B},
  volume={110},
  number={11},
  pages={115114},
  year={2024},
  publisher={APS}
}

@article{stormer1999fractional,
  title = {The fractional quantum Hall effect},
  author = {Stormer, Horst L. and Tsui, Daniel C. and Gossard, Arthur C.},
  journal = {Rev. Mod. Phys.},
  volume = {71},
  issue = {2},
  pages = {S298--S305},
  numpages = {0},
  year = {1999},
  month = {Mar},
  publisher = {American Physical Society},
  doi = {10.1103/RevModPhys.71.S298},
  url = {https://link.aps.org/doi/10.1103/RevModPhys.71.S298}
}

@article{regnault2011fractional,
  title = {Fractional Chern Insulator},
  author = {Regnault, N. and Bernevig, B. Andrei},
  journal = {Phys. Rev. X},
  volume = {1},
  issue = {2},
  pages = {021014},
  numpages = {14},
  year = {2011},
  month = {Dec},
  publisher = {American Physical Society},
  doi = {10.1103/PhysRevX.1.021014},
  url = {https://link.aps.org/doi/10.1103/PhysRevX.1.021014}
}

@article{sheng2011fractional,
  title={Fractional quantum Hall effect in the absence of Landau levels},
  author={Sheng, DN and Gu, Zheng-Cheng and Sun, Kai and Sheng, L},
  journal={Nature communications},
  volume={2},
  number={1},
  pages={389},
  year={2011},
  publisher={Nature Publishing Group UK London}
}

@article{neupert2011fractional,
  title = {Fractional Quantum Hall States at Zero Magnetic Field},
  author = {Neupert, Titus and Santos, Luiz and Chamon, Claudio and Mudry, Christopher},
  journal = {Phys. Rev. Lett.},
  volume = {106},
  issue = {23},
  pages = {236804},
  numpages = {4},
  year = {2011},
  month = {Jun},
  publisher = {American Physical Society},
  doi = {10.1103/PhysRevLett.106.236804},
  url = {https://link.aps.org/doi/10.1103/PhysRevLett.106.236804}
}

@article{tang2011high,
  title = {High-Temperature Fractional Quantum Hall States},
  author = {Tang, Evelyn and Mei, Jia-Wei and Wen, Xiao-Gang},
  journal = {Phys. Rev. Lett.},
  volume = {106},
  issue = {23},
  pages = {236802},
  numpages = {4},
  year = {2011},
  month = {Jun},
  publisher = {American Physical Society},
  doi = {10.1103/PhysRevLett.106.236802},
  url = {https://link.aps.org/doi/10.1103/PhysRevLett.106.236802}
}

@article{sun2011nearly,
  title = {Nearly Flatbands with Nontrivial Topology},
  author = {Sun, Kai and Gu, Zhengcheng and Katsura, Hosho and Das Sarma, S.},
  journal = {Phys. Rev. Lett.},
  volume = {106},
  issue = {23},
  pages = {236803},
  numpages = {4},
  year = {2011},
  month = {Jun},
  publisher = {American Physical Society},
  doi = {10.1103/PhysRevLett.106.236803},
  url = {https://link.aps.org/doi/10.1103/PhysRevLett.106.236803}
}

@article{cai2023signatures,
  title={{Signatures of fractional quantum anomalous Hall states in twisted MoTe$_2$}},
  author={Cai, Jiaqi and Anderson, Eric and Wang, Chong and Zhang, Xiaowei and Liu, Xiaoyu and Holtzmann, William and Zhang, Yinong and Fan, Fengren and Taniguchi, Takashi and Watanabe, Kenji and others},
  journal={Nature},
  volume={622},
  number={7981},
  pages={63--68},
  year={2023},
  publisher={Nature Publishing Group UK London}
}

@article{park2023observation,
  title={Observation of fractionally quantized anomalous Hall effect},
  author={Park, Heonjoon and Cai, Jiaqi and Anderson, Eric and Zhang, Yinong and Zhu, Jiayi and Liu, Xiaoyu and Wang, Chong and Holtzmann, William and Hu, Chaowei and Liu, Zhaoyu and others},
  journal={Nature},
  volume={622},
  number={7981},
  pages={74--79},
  year={2023},
  publisher={Nature Publishing Group UK London}
}

@article{zeng2023thermodynamic,
  title={{Thermodynamic evidence of fractional Chern insulator in moir{\'e} MoTe$_2$}},
  author={Zeng, Yihang and Xia, Zhengchao and Kang, Kaifei and Zhu, Jiacheng and Kn{\"u}ppel, Patrick and Vaswani, Chirag and Watanabe, Kenji and Taniguchi, Takashi and Mak, Kin Fai and Shan, Jie},
  journal={Nature},
  volume={622},
  number={7981},
  pages={69--73},
  year={2023},
  publisher={Nature Publishing Group UK London}
}

@article{xu2023observation,
  title = {{Observation of Integer and Fractional Quantum Anomalous Hall Effects in Twisted Bilayer ${\mathrm{MoTe}}_{2}$}},
  author = {Xu, Fan and Sun, Zheng and Jia, Tongtong and Liu, Chang and Xu, Cheng and Li, Chushan and Gu, Yu and Watanabe, Kenji and Taniguchi, Takashi and Tong, Bingbing and Jia, Jinfeng and Shi, Zhiwen and Jiang, Shengwei and Zhang, Yang and Liu, Xiaoxue and Li, Tingxin},
  journal = {Phys. Rev. X},
  volume = {13},
  issue = {3},
  pages = {031037},
  numpages = {12},
  year = {2023},
  month = {Sep},
  publisher = {American Physical Society},
  doi = {10.1103/PhysRevX.13.031037},
  url = {https://link.aps.org/doi/10.1103/PhysRevX.13.031037}
}

@article{lu2024fractional,
  title={Fractional quantum anomalous Hall effect in multilayer graphene},
  author={Lu, Zhengguang and Han, Tonghang and Yao, Yuxuan and Reddy, Aidan P and Yang, Jixiang and Seo, Junseok and Watanabe, Kenji and Taniguchi, Takashi and Fu, Liang and Ju, Long},
  journal={Nature},
  volume={626},
  number={8000},
  pages={759--764},
  year={2024},
  publisher={Nature Publishing Group UK London}
}

@article{lu2025extended,
  title={Extended quantum anomalous Hall states in graphene/hBN moir{\'e} superlattices},
  author={Lu, Zhengguang and Han, Tonghang and Yao, Yuxuan and Hadjri, Zach and Yang, Jixiang and Seo, Junseok and Shi, Lihan and Ye, Shenyong and Watanabe, Kenji and Taniguchi, Takashi and others},
  journal={Nature},
  volume={637},
  number={8048},
  pages={1090--1095},
  year={2025},
  publisher={Nature Publishing Group UK London}
}

@article{waters2025chern,
  title = {Chern Insulators at Integer and Fractional Filling in Moir\'e Pentalayer Graphene},
  author = {Waters, Dacen and Okounkova, Anna and Su, Ruiheng and Zhou, Boran and Yao, Jiang and Watanabe, Kenji and Taniguchi, Takashi and Xu, Xiaodong and Zhang, Ya-Hui and Folk, Joshua and Yankowitz, Matthew},
  journal = {Phys. Rev. X},
  volume = {15},
  issue = {1},
  pages = {011045},
  numpages = {8},
  year = {2025},
  month = {Feb},
  publisher = {American Physical Society},
  doi = {10.1103/PhysRevX.15.011045},
  url = {https://link.aps.org/doi/10.1103/PhysRevX.15.011045}
}

@article{xie2025tunable,
  title={Tunable fractional Chern insulators in rhombohedral graphene superlattices},
  author={Xie, Jian and Huo, Zihao and Lu, Xin and Feng, Zuo and Zhang, Zaizhe and Wang, Wenxuan and Yang, Qiu and Watanabe, Kenji and Taniguchi, Takashi and Liu, Kaihui and others},
  journal={Nature Materials},
  volume={24},
  number={7},
  pages={1042--1048},
  year={2025},
  publisher={Nature Publishing Group UK London}
}

@article{aronson2025displacement,
  title = {{Displacement Field-Controlled Fractional Chern Insulators and Charge Density Waves in a Graphene/hBN Moir\'e Superlattice}},
  author = {Aronson, Samuel H. and Han, Tonghang and Lu, Zhengguang and Yao, Yuxuan and Butler, Jackson P. and Watanabe, Kenji and Taniguchi, Takashi and Ju, Long and Ashoori, Raymond C.},
  journal = {Phys. Rev. X},
  volume = {15},
  issue = {3},
  pages = {031026},
  numpages = {9},
  year = {2025},
  month = {Jul},
  publisher = {American Physical Society},
  doi = {10.1103/75gl-jzl6},
  url = {https://link.aps.org/doi/10.1103/75gl-jzl6}
}

@article{choi2025superconductivity,
  title={Superconductivity and quantized anomalous Hall effect in rhombohedral graphene},
  author={Choi, Youngjoon and Choi, Ysun and Valentini, Marco and Patterson, Caitlin L and Holleis, Ludwig FW and Sheekey, Owen I and Stoyanov, Hari and Cheng, Xiang and Taniguchi, Takashi and Watanabe, Kenji and others},
  journal={Nature},
  volume={639},
  number={8054},
  pages={342--347},
  year={2025},
  publisher={Nature Publishing Group UK London}
}

@article{butler2026fci,
  title={1/3 Fractional and Gapless Integer Quantum Anomalous Hall States in Rhombohedral Graphene},
  author={Butler, Jackson P and Han, Tonghang and DiFabbio, Andrew and Hadjri, Zach and Aitken, Emily and Watanabe, Kenji and Taniguchi, Takashi and Ju, Long and Ashoori, Raymond C},
  journal={arXiv preprint arXiv:2606.06450},
  year={2026}
}

@article{dong2024anomalous,
  title = {Anomalous Hall Crystals in Rhombohedral Multilayer Graphene. I. Interaction-Driven Chern Bands and Fractional Quantum Hall States at Zero Magnetic Field},
  author = {Dong, Junkai and Wang, Taige and Wang, Tianle and Soejima, Tomohiro and Zaletel, Michael P. and Vishwanath, Ashvin and Parker, Daniel E.},
  journal = {Phys. Rev. Lett.},
  volume = {133},
  issue = {20},
  pages = {206503},
  numpages = {11},
  year = {2024},
  month = {Nov},
  publisher = {American Physical Society},
  doi = {10.1103/PhysRevLett.133.206503},
  url = {https://link.aps.org/doi/10.1103/PhysRevLett.133.206503}
}

@article{dong2024theory,
  title = {Theory of Quantum Anomalous Hall Phases in Pentalayer Rhombohedral Graphene Moir\'e Structures},
  author = {Dong, Zhihuan and Patri, Adarsh S. and Senthil, T.},
  journal = {Phys. Rev. Lett.},
  volume = {133},
  issue = {20},
  pages = {206502},
  numpages = {6},
  year = {2024},
  month = {Nov},
  publisher = {American Physical Society},
  doi = {10.1103/PhysRevLett.133.206502},
  url = {https://link.aps.org/doi/10.1103/PhysRevLett.133.206502}
}

@article{zhou2024fractional,
  title = {Fractional Quantum Anomalous Hall Effect in Rhombohedral Multilayer Graphene in the Moir\'eless Limit},
  author = {Zhou, Boran and Yang, Hui and Zhang, Ya-Hui},
  journal = {Phys. Rev. Lett.},
  volume = {133},
  issue = {20},
  pages = {206504},
  numpages = {7},
  year = {2024},
  month = {Nov},
  publisher = {American Physical Society},
  doi = {10.1103/PhysRevLett.133.206504},
  url = {https://link.aps.org/doi/10.1103/PhysRevLett.133.206504}
}

@article{guo2024fractional,
  title = {Fractional Chern insulator states in multilayer graphene moir\'e superlattices},
  author = {Guo, Zhongqing and Lu, Xin and Xie, Bo and Liu, Jianpeng},
  journal = {Phys. Rev. B},
  volume = {110},
  issue = {7},
  pages = {075109},
  numpages = {13},
  year = {2024},
  month = {Aug},
  publisher = {American Physical Society},
  doi = {10.1103/PhysRevB.110.075109},
  url = {https://link.aps.org/doi/10.1103/PhysRevB.110.075109}
}

@article{kwan2025moire,
  title = {{Moir\'e fractional Chern insulators. III. Hartree-Fock phase diagram, magic angle regime for Chern insulator states, role of moir\'e potential, and Goldstone gaps in rhombohedral graphene superlattices}},
  author = {Kwan, Yves H. and Yu, Jiabin and Herzog-Arbeitman, Jonah and Efetov, Dmitri K. and Regnault, Nicolas and Bernevig, B. Andrei},
  journal = {Phys. Rev. B},
  volume = {112},
  issue = {7},
  pages = {075109},
  numpages = {22},
  year = {2025},
  month = {Aug},
  publisher = {American Physical Society},
  doi = {10.1103/PhysRevB.112.075109},
  url = {https://link.aps.org/doi/10.1103/PhysRevB.112.075109}
}

@article{zeng2024sublattice,
  title = {Sublattice Structure and Topology in Spontaneously Crystallized Electronic States},
  author = {Zeng, Yongxin and Guerci, Daniele and Cr\'epel, Valentin and Millis, Andrew J. and Cano, Jennifer},
  journal = {Phys. Rev. Lett.},
  volume = {132},
  issue = {23},
  pages = {236601},
  numpages = {8},
  year = {2024},
  month = {Jun},
  publisher = {American Physical Society},
  doi = {10.1103/PhysRevLett.132.236601},
  url = {https://link.aps.org/doi/10.1103/PhysRevLett.132.236601}
}

@article{tan2024parent,
  title = {Parent Berry Curvature and the Ideal Anomalous Hall Crystal},
  author = {Tan, Tixuan and Devakul, Trithep},
  journal = {Phys. Rev. X},
  volume = {14},
  issue = {4},
  pages = {041040},
  numpages = {15},
  year = {2024},
  month = {Nov},
  publisher = {American Physical Society},
  doi = {10.1103/PhysRevX.14.041040},
  url = {https://link.aps.org/doi/10.1103/PhysRevX.14.041040}
}

@article{soejima2024anomalous,
  title = {Anomalous Hall crystals in rhombohedral multilayer graphene. II. General mechanism and a minimal model},
  author = {Soejima, Tomohiro and Dong, Junkai and Wang, Taige and Wang, Tianle and Zaletel, Michael P. and Vishwanath, Ashvin and Parker, Daniel E.},
  journal = {Phys. Rev. B},
  volume = {110},
  issue = {20},
  pages = {205124},
  numpages = {20},
  year = {2024},
  month = {Nov},
  publisher = {American Physical Society},
  doi = {10.1103/PhysRevB.110.205124},
  url = {https://link.aps.org/doi/10.1103/PhysRevB.110.205124}
}

@article{dong2024stability,
  title = {Stability of anomalous Hall crystals in multilayer rhombohedral graphene},
  author = {Dong, Zhihuan and Patri, Adarsh S. and Senthil, T.},
  journal = {Phys. Rev. B},
  volume = {110},
  issue = {20},
  pages = {205130},
  numpages = {24},
  year = {2024},
  month = {Nov},
  publisher = {American Physical Society},
  doi = {10.1103/PhysRevB.110.205130},
  url = {https://link.aps.org/doi/10.1103/PhysRevB.110.205130}
}

@article{zhou2025new,
  title = {New Classes of Quantum Anomalous Hall Crystals in Multilayer Graphene},
  author = {Zhou, Boran and Zhang, Ya-Hui},
  journal = {Phys. Rev. Lett.},
  volume = {135},
  issue = {3},
  pages = {036501},
  numpages = {6},
  year = {2025},
  month = {Jul},
  publisher = {American Physical Society},
  doi = {10.1103/26pl-gkh7},
  url = {https://link.aps.org/doi/10.1103/26pl-gkh7}
}

@article{tan2025variational,
  title = {Variational Wave-Function Analysis of the Fractional Anomalous Hall Crystal},
  author = {Tan, Tixuan and May-Mann, Julian and Devakul, Trithep},
  journal = {Phys. Rev. Lett.},
  volume = {135},
  issue = {3},
  pages = {036604},
  numpages = {8},
  year = {2025},
  month = {Jul},
  publisher = {American Physical Society},
  doi = {10.1103/dd2d-kk3w},
  url = {https://link.aps.org/doi/10.1103/dd2d-kk3w}
}

@article{desrochers2026energetics,
  title={Energetics of fractional anomalous Hall crystals in rhombohedral graphene},
  author={Desrochers, F{\'e}lix and Vishwanath, Ashvin},
  journal={arXiv preprint arXiv:2607.08822},
  year={2026}
}

@article{maymann2026skyrmion,
  title={Skyrmion Fractional Chern Insulator: An Intrinsically Multiband Route to Fractionalization in Rhombohedral Graphene},
  author={May-Mann, Julian and Tan, Tixuan and Ledwith, Patrick J and Shi, Zhengyan Darius and Devakul, Trithep},
  journal={arXiv preprint arXiv:2608.14535},
  year={2026}
}

@article{regnault2026moire,
  title={The ``Moir\'e Capacitor Effect" and Stabilization of Fractional Chern Insulators in Rhombohedral Graphene Superlattices},
  author={Regnault, Nicolas and Li, Heqiu and Kwan, Yves H and Bernevig, B Andrei and Herzog-Arbeitman, Jonah},
  journal={arXiv preprint arXiv:2608.12452},
  year={2026}
}

@article{park2023topological,
  title = {Topological flat bands in rhombohedral tetralayer and multilayer graphene on hexagonal boron nitride moir\'e superlattices},
  author = {Park, Youngju and Kim, Yeonju and Chittari, Bheema Lingam and Jung, Jeil},
  journal = {Phys. Rev. B},
  volume = {108},
  issue = {15},
  pages = {155406},
  numpages = {16},
  year = {2023},
  month = {Oct},
  publisher = {American Physical Society},
  doi = {10.1103/PhysRevB.108.155406},
  url = {https://link.aps.org/doi/10.1103/PhysRevB.108.155406}
}

@article{herzog2024moire,
  title = {Moir\'e fractional Chern insulators. II. First-principles calculations and continuum models of rhombohedral graphene superlattices},
  author = {Herzog-Arbeitman, Jonah and Wang, Yuzhi and Liu, Jiaxuan and Tam, Pok Man and Qi, Ziyue and Jia, Yujin and Efetov, Dmitri K. and Vafek, Oskar and Regnault, Nicolas and Weng, Hongming and Wu, Quansheng and Bernevig, B. Andrei and Yu, Jiabin},
  journal = {Phys. Rev. B},
  volume = {109},
  issue = {20},
  pages = {205122},
  numpages = {46},
  year = {2024},
  month = {May},
  publisher = {American Physical Society},
  doi = {10.1103/PhysRevB.109.205122},
  url = {https://link.aps.org/doi/10.1103/PhysRevB.109.205122}
}

@article{yu2025moire,
  title = {{Moir\'e fractional Chern insulators. IV. Fluctuation-driven collapse in multiband exact diagonalization calculations on rhombohedral graphene}},
  author = {Yu, Jiabin and Herzog-Arbeitman, Jonah and Kwan, Yves H. and Regnault, Nicolas and Bernevig, B. Andrei},
  journal = {Phys. Rev. B},
  volume = {112},
  issue = {7},
  pages = {075110},
  numpages = {73},
  year = {2025},
  month = {Aug},
  publisher = {American Physical Society},
  doi = {10.1103/PhysRevB.112.075110},
  url = {https://link.aps.org/doi/10.1103/PhysRevB.112.075110}
}

@article{bernevig2025berry,
  title={``Berry Trashcan" Model of Interacting Electrons in Rhombohedral Graphene},
  author={Bernevig, B Andrei and Kwan, Yves H},
  journal={arXiv preprint arXiv:2503.09692},
  year={2025}
}

@article{li2025multi,
  title={Multi-Band Exact Diagonalization and an Iteration Approach to Hunt For Fractional Chern Insulators in Rhombohedral Multilayer Graphene},
  author={Li, Heqiu and Bernevig, B Andrei and Regnault, Nicolas},
  journal={arXiv preprint arXiv:2504.20140},
  year={2025}
}

@article{han2025signatures,
  title={Signatures of chiral superconductivity in rhombohedral graphene},
  author={Han, Tonghang and Lu, Zhengguang and Hadjri, Zach and Shi, Lihan and Wu, Zhenghan and Xu, Wei and Yao, Yuxuan and Cotten, Armel A and Sedeh, Omid Sharifi and Weldeyesus, Henok and others},
  journal={Nature},
  pages={1--3},
  year={2025},
  publisher={Nature Publishing Group UK London}
}

@article{seo2025family,
  title={Family of Unconventional Superconductivities in Crystalline Graphene},
  author={Seo, Junseok and Cotten, Armel A and Xu, Mingchi and Sedeh, Omid Sharifi and Weldeyesus, Henok and Han, Tonghang and Lu, Zhengguang and Wu, Zhenghan and Ye, Shenyong and Xu, Wei and others},
  journal={arXiv preprint arXiv:2509.03295},
  year={2025}
}

@article{han2026evidence,
  title={Evidence of Metallic Wigner Crystal in Rhombohedral Graphene},
  author={Han, Tonghang and Butler, Jackson P and Ye, Shenyong and Hua, Zhenqi and Dutta, Surajit and Hadjri, Zach and Wu, Zhenghan and Yang, Jixiang and Seo, Junseok and Pattanakanvijit, Phatthanon and others},
  journal={arXiv preprint arXiv:2604.00113},
  year={2026}
}

@article{jain1989composite,
  title = {Composite-fermion approach for the fractional quantum Hall effect},
  author = {Jain, J. K.},
  journal = {Phys. Rev. Lett.},
  volume = {63},
  issue = {2},
  pages = {199--202},
  numpages = {0},
  year = {1989},
  month = {Jul},
  publisher = {American Physical Society},
  doi = {10.1103/PhysRevLett.63.199},
  url = {https://link.aps.org/doi/10.1103/PhysRevLett.63.199}
}

@book{jain2007composite,
  title={Composite fermions},
  author={Jain, Jainendra K},
  year={2007},
  publisher={Cambridge University Press}
}

@article{jain2015composite,
  title={Composite fermion theory of exotic fractional quantum Hall effect},
  author={Jain, Jainendra K},
  journal={Annu. Rev. Condens. Matter Phys.},
  volume={6},
  number={1},
  pages={39--62},
  year={2015},
  publisher={Annual Reviews}
}

@article{halperin1993theory,
  title = {Theory of the half-filled Landau level},
  author = {Halperin, B. I. and Lee, Patrick A. and Read, Nicholas},
  journal = {Phys. Rev. B},
  volume = {47},
  issue = {12},
  pages = {7312--7343},
  numpages = {0},
  year = {1993},
  month = {Mar},
  publisher = {American Physical Society},
  doi = {10.1103/PhysRevB.47.7312},
  url = {https://link.aps.org/doi/10.1103/PhysRevB.47.7312}
}

@article{lopez1991fractional,
  title = {Fractional quantum Hall effect and Chern-Simons gauge theories},
  author = {Lopez, Ana and Fradkin, Eduardo},
  journal = {Phys. Rev. B},
  volume = {44},
  issue = {10},
  pages = {5246--5262},
  numpages = {0},
  year = {1991},
  month = {Sep},
  publisher = {American Physical Society},
  doi = {10.1103/PhysRevB.44.5246},
  url = {https://link.aps.org/doi/10.1103/PhysRevB.44.5246}
}

@article{sohal2018cf,
  title = {Chern-Simons composite fermion theory of fractional Chern insulators},
  author = {Sohal, Ramanjit and Santos, Luiz H. and Fradkin, Eduardo},
  journal = {Phys. Rev. B},
  volume = {97},
  issue = {12},
  pages = {125131},
  numpages = {12},
  year = {2018},
  month = {Mar},
  publisher = {American Physical Society},
  doi = {10.1103/PhysRevB.97.125131},
  url = {https://link.aps.org/doi/10.1103/PhysRevB.97.125131}
}

@article{lu2024CF,
  title = {{Fractional Chern Insulators in Twisted Bilayer ${\mathrm{MoTe}}_{2}$: A Composite Fermion Perspective}},
  author = {Lu, Tianhong and Santos, Luiz H.},
  journal = {Phys. Rev. Lett.},
  volume = {133},
  issue = {18},
  pages = {186602},
  numpages = {6},
  year = {2024},
  month = {Oct},
  publisher = {American Physical Society},
  doi = {10.1103/PhysRevLett.133.186602},
  url = {https://link.aps.org/doi/10.1103/PhysRevLett.133.186602}
}

@article{gassner2025flux,
  title = {Flux Attachment Theory of Fractional Excitonic Insulators},
  author = {Gassner, Steven and Stern, Ady and Kane, C. L.},
  journal = {Phys. Rev. Lett.},
  volume = {135},
  issue = {18},
  pages = {186602},
  numpages = {8},
  year = {2025},
  month = {Oct},
  publisher = {American Physical Society},
  doi = {10.1103/x789-kxy3},
  url = {https://link.aps.org/doi/10.1103/x789-kxy3}
}

@article{hu2026cf,
  title = {Composite Fermion Theory of Fractional Chern Insulator Stability},
  author = {Hu, Xiaodong and Ran, Ying and Xiao, Di},
  journal = {Phys. Rev. Lett.},
  volume = {136},
  issue = {6},
  pages = {066504},
  numpages = {8},
  year = {2026},
  month = {Feb},
  publisher = {American Physical Society},
  doi = {10.1103/shrz-b692},
  url = {https://link.aps.org/doi/10.1103/shrz-b692}
}

@article{jin2026cf,
  title = {Generalizing the composite fermion theory for fractional Chern insulators},
  author = {Jin, Hao and Shi, Junren},
  journal = {Phys. Rev. B},
  volume = {114},
  issue = {11},
  pages = {L111106},
  numpages = {6},
  year = {2026},
  month = {Aug},
  publisher = {American Physical Society},
  doi = {10.1103/zsxs-mcxc},
  url = {https://link.aps.org/doi/10.1103/zsxs-mcxc}
}

@article{xiao2005berry,
  title = {Berry Phase Correction to Electron Density of States in Solids},
  author = {Xiao, Di and Shi, Junren and Niu, Qian},
  journal = {Phys. Rev. Lett.},
  volume = {95},
  issue = {13},
  pages = {137204},
  numpages = {4},
  year = {2005},
  month = {Sep},
  publisher = {American Physical Society},
  doi = {10.1103/PhysRevLett.95.137204},
  url = {https://link.aps.org/doi/10.1103/PhysRevLett.95.137204}
}

@article{xiao2010berry,
  title = {Berry phase effects on electronic properties},
  author = {Xiao, Di and Chang, Ming-Che and Niu, Qian},
  journal = {Rev. Mod. Phys.},
  volume = {82},
  issue = {3},
  pages = {1959--2007},
  numpages = {0},
  year = {2010},
  month = {Jul},
  publisher = {American Physical Society},
  doi = {10.1103/RevModPhys.82.1959},
  url = {https://link.aps.org/doi/10.1103/RevModPhys.82.1959}
}

@article{zou2025valley,
  title={Valley Order in Moir$\backslash$'e Topological Insulators},
  author={Zou, Bo and Li, Anzhuoer and MacDonald, Allan H},
  journal={arXiv preprint arXiv:2509.07784},
  year={2025}
}

@article{murthy2003hamiltonian,
  title={Hamiltonian theories of the fractional quantum Hall effect},
  author={Murthy, Ganpathy and Shankar, R},
  journal={Reviews of Modern Physics},
  volume={75},
  number={4},
  pages={1101},
  year={2003},
  publisher={APS}
}

@article{wang2021exact,
  title = {Exact Landau Level Description of Geometry and Interaction in a Flatband},
  author = {Wang, Jie and Cano, Jennifer and Millis, Andrew J. and Liu, Zhao and Yang, Bo},
  journal = {Phys. Rev. Lett.},
  volume = {127},
  issue = {24},
  pages = {246403},
  numpages = {6},
  year = {2021},
  month = {Dec},
  publisher = {American Physical Society},
  doi = {10.1103/PhysRevLett.127.246403},
  url = {https://link.aps.org/doi/10.1103/PhysRevLett.127.246403}
}

@article{roy2014band,
  title = {Band geometry of fractional topological insulators},
  author = {Roy, Rahul},
  journal = {Phys. Rev. B},
  volume = {90},
  issue = {16},
  pages = {165139},
  numpages = {7},
  year = {2014},
  month = {Oct},
  publisher = {American Physical Society},
  doi = {10.1103/PhysRevB.90.165139},
  url = {https://link.aps.org/doi/10.1103/PhysRevB.90.165139}
}

@article{claassen2015position,
  title = {Position-Momentum Duality and Fractional Quantum Hall Effect in Chern Insulators},
  author = {Claassen, Martin and Lee, Ching Hua and Thomale, Ronny and Qi, Xiao-Liang and Devereaux, Thomas P.},
  journal = {Phys. Rev. Lett.},
  volume = {114},
  issue = {23},
  pages = {236802},
  numpages = {6},
  year = {2015},
  month = {Jun},
  publisher = {American Physical Society},
  doi = {10.1103/PhysRevLett.114.236802},
  url = {https://link.aps.org/doi/10.1103/PhysRevLett.114.236802}
}

@article{ledwith2023vortexability,
  title = {Vortexability: A unifying criterion for ideal fractional Chern insulators},
  author = {Ledwith, Patrick J. and Vishwanath, Ashvin and Parker, Daniel E.},
  journal = {Phys. Rev. B},
  volume = {108},
  issue = {20},
  pages = {205144},
  numpages = {20},
  year = {2023},
  month = {Nov},
  publisher = {American Physical Society},
  doi = {10.1103/PhysRevB.108.205144},
  url = {https://link.aps.org/doi/10.1103/PhysRevB.108.205144}
}

@article{jackson2015geometric,
  title={Geometric stability of topological lattice phases},
  author={Jackson, Thomas S and M{\"o}ller, Gunnar and Roy, Rahul},
  journal={Nature communications},
  volume={6},
  number={1},
  pages={8629},
  year={2015},
  publisher={Nature Publishing Group UK London}
}

@article{moore1991nonabelions,
  title={Nonabelions in the fractional quantum Hall effect},
  author={Moore, Gregory and Read, Nicholas},
  journal={Nuclear Physics B},
  volume={360},
  number={2-3},
  pages={362--396},
  year={1991},
  publisher={Elsevier}
}


\onecolumngrid
\newpage
\makeatletter 

\begin{center}
\textbf{\large Supplemental Material for ``\@title ''} \\[10pt]
\end{center}
\vspace{20pt}

\setcounter{figure}{0}
\setcounter{section}{0}
\setcounter{equation}{0}

\renewcommand{\thefigure}{S\@arabic\c@figure}
\renewcommand{\theequation}{S\@arabic\c@equation}
\renewcommand{\theHfigure}{suppfig.\arabic{figure}}
\renewcommand{\theHequation}{suppeq.\arabic{equation}}
\makeatother



\section{Rhombohedral pentalayer graphene (R5G)}
The single-particle Hamiltonian of R5G is a $10\times 10$ matrix acting in layer-sublattice space. Written in layer space, it takes the form
\begin{equation} \label{eq:H_R5G_supp}
H_{\mathrm{R5G}} = \begin{pmatrix}
h_0+2u_D & h_1 & h_2 & 0 & 0 \\
h_1^{\dagger} & h_0+u_D & h_1 & h_2 & 0 \\
h_2^{\dagger} & h_1^{\dagger} & h_0 & h_1 & h_2 \\
0 & h_2^{\dagger} & h_1^{\dagger} & h_0-u_D & h_1 \\
0 & 0 & h_2^{\dagger} & h_1^{\dagger} & h_0-2u_D
\end{pmatrix},
\end{equation}
where each $h$ is a $2\times 2$ matrix in sublattice space. The diagonal blocks
\begin{equation}
h_0(\bm p) = \begin{pmatrix}
0 & v_0 p_- \\
v_0 p_+ & 0
\end{pmatrix},
\end{equation}
where $p_{\pm} = \chi_v p_x \pm ip_y$ with $\chi_v=\pm1$ depending on the valley. We choose $\chi_v=-1$, the valley whose conduction-band Berry curvature is positive in our convention. This choice is also the one for which the Landau-orbital indices increase along the layer--sublattice sequence in Eqs.~\eqref{eq:psi_LLL} and \eqref{eq:supp_LLL_orbital_content}. The off-diagonal blocks
\begin{equation}
h_1(\bm p) = \begin{pmatrix}
v_4 p_- & v_3 p_+ \\
\gamma_1 & v_4 p_-
\end{pmatrix}, \quad
h_2 = \begin{pmatrix}
0 & \gamma_2/2 \\
0 & 0
\end{pmatrix}
\end{equation}
describe hopping between neighboring and next-neighboring layers, respectively. We model the applied displacement field as a potential difference $u_D$ between adjacent layers. The numerical parameters are taken from Ref.~\cite{zhou2021half}, where they were chosen to match experimental data in rhombohedral trilayer graphene: $v_0 = \SI{1.0e6}{m/s}$, $\gamma_1 = \SI{380}{meV}$, $\gamma_2 = \SI{-15}{meV}$, $v_3 = \SI{-9.4e4}{m/s}$, $v_4 = \SI{-4.6e4}{m/s}$.

\section{Flux attachment}

We start from a general interacting Hamiltonian for two-dimensional electrons with internal orbital structure.  
For rhombohedral pentalayer graphene (R5G), the electron operator is a ten-component spinor in layer--sublattice space,
\begin{equation}
\psi(\bm r) = (\psi_{A_1}, \psi_{B_1}, \psi_{A_2}, \psi_{B_2}, \dots, \psi_{A_5}, \psi_{B_5})^{\top}.
\label{eq:psi_spinor}
\end{equation}
The single-particle part contains the kinetic Hamiltonian $H_0(-i\nabla)$ [Eq.~\eqref{eq:H_R5G_supp}] and the moiré potential $V_{\mathrm{M}}(\bm r)$. The full interacting Hamiltonian is
\begin{equation}
H = \int d^2 r\, \psi^\dagger(\bm r) [H_0(-i\nabla) + V_{\mathrm{M}}(\bm r)] \psi(\bm r)
+ \frac12 \int d^2 r d^2 r'\, V(\bm r-\bm r')\, 
\psi^\dagger(\bm r)\psi^\dagger(\bm r')\psi(\bm r')\psi(\bm r),
\label{eq:H_full}
\end{equation}
where $V(\bm r-\bm r')$ represents the Coulomb interaction.

To attach $2s$ flux quanta to each active conduction-band electron, we perform a singular gauge transformation using the unitary operator
\begin{equation}
U_{\chi_{\rm FA}} = \exp\Big[-i 2s\chi_{\rm FA} \sum_{i<j} \arg(\bm r_i - \bm r_j)\Big],
\label{eq:U_fluxattach}
\end{equation}
where $\arg(\bm r_i-\bm r_j)$ is the polar angle of the vector connecting two electrons and $\chi_{\rm FA}=\pm1$ specifies the chirality of the attached flux.
The transformed fermion operator is
\begin{equation}
\tilde{\psi}(\bm r) = U_{\chi_{\rm FA}}^\dagger \psi(\bm r) U_{\chi_{\rm FA}}
= e^{i\Lambda(\bm r)} \psi(\bm r),
\qquad
\Lambda(\bm r)=2s\chi_{\rm FA}\int d^2 r'\,\arg(\bm r-\bm r')n(\bm r'),
\qquad 
n(\bm r) = \psi^\dagger(\bm r) \psi(\bm r).
\label{eq:psi_tilde}
\end{equation}
In Eqs.~\eqref{eq:U_fluxattach}--\eqref{eq:S_CS}, $\psi$ and $n$ denote the active conduction-band field and its density; the ``act'' subscript is suppressed. The filled valence bands are treated as an inert background and are subtracted from $n$, as stated in the main text. At mean field, the statistical vector potential obtained from this active density is embedded in the full ten-component continuum Hamiltonian. This preserves the layer--sublattice structure of the conduction-band Landau level without attaching flux to the inert bands. The transformation attaches $2s$ flux quanta to each active electron and defines the composite-fermion operators $\tilde{\psi}$.
The position-dependent phase introduces an emergent gauge field whenever spatial derivatives act on $\psi$:
\begin{equation}
\nabla \psi(\bm r) = e^{-i\Lambda(\bm r)}
\big[\nabla - i \bm a(\bm r)\big] \tilde{\psi}(\bm r),
\qquad
\bm a(\bm r) = \nabla\Lambda(\bm r)
=2s\chi_{\rm FA} \int d^2 r' \frac{\hat{\bm z}\times(\bm r-\bm r')}{|\bm r-\bm r'|^2} n(\bm r').
\label{eq:a_field}
\end{equation}
Thus $-i\nabla\psi=e^{-i\Lambda}(-i\nabla-\bm a)\tilde\psi$. The field $\bm a$ has units of inverse length. We define the physical statistical vector potential by
$\bm A^*=-(\hbar/e)\bm a$, so the attached magnetic field is
\begin{equation}
\Delta B(\bm r)
=-\frac{\hbar}{e}\nabla \times \bm a(\bm r)
=
-\frac{4\pi s\chi_{\rm FA}\hbar}{e} n(\bm r),
\label{eq:flux_attachment_constraint}
\end{equation}
which enforces the flux-attachment condition. We choose $\chi_{\rm FA}=+1$ throughout, so that $\Delta B<0$ for electron doping.
In terms of composite fermions, the transformed Hamiltonian becomes
\begin{equation}
H = \int d^2 r\, \tilde{\psi}^\dagger(\bm r) [H_0(-i\nabla - \bm a(\bm r)) + V_{\mathrm{M}}(\bm r)] \tilde{\psi}(\bm r)
+ \frac12 \int d^2 r d^2 r'\, V(\bm r-\bm r') n(\bm r) n(\bm r').
\label{eq:H_CF}
\end{equation}

The Hamiltonian in Eq.~\eqref{eq:H_CF} should be supplemented by the
operator constraint in Eq.~\eqref{eq:flux_attachment_constraint}.
Equivalently, one may promote $\bm a$ to an independent emergent gauge field
$a_\mu=(a_0,\bm a)$ in a path-integral formulation. The flux-attachment
constraint is then imposed locally by the matter coupling and Chern--Simons action
\begin{equation}
S[a_0,\bm a;\tilde\psi]\supset
-\int dt\,d^2r\,a_0 n
+S_{\rm CS},
\qquad
S_{\rm CS}
=
\frac{\chi_{\rm FA}}{8\pi s}
\int dt\,d^2r\,
\epsilon^{\mu\nu\lambda}a_\mu\partial_\nu a_\lambda .
\label{eq:S_CS}
\end{equation}
Varying the action with respect to $a_0$ gives
\begin{equation}
\frac{\chi_{\rm FA}}{4\pi s}\nabla\times\bm a(\bm r,t)-n(\bm r,t)=0,
\end{equation}
which reproduces Eq.~\eqref{eq:flux_attachment_constraint}.

In the mean-field approximation, the statistical gauge field is replaced by its spatial average.  
For a uniform electron density $\bar n$, the attached magnetic field becomes
\begin{equation}
\Delta B = \nabla \times \mathbf A^* = -\frac{4\pi s\chi_{\rm FA}\hbar}{e}\bar n,
\label{eq:b_mean}
\end{equation}
where $\mathbf A^*$ is the mean-field statistical vector potential.  
Temporarily neglecting interactions, the resulting mean-field Hamiltonian reads
\begin{equation}
H_{\rm MF} = H_0\!\left(-i\nabla + \frac{e}{\hbar}\mathbf A^*(\bm r)\right) + V_{\mathrm{M}}(\bm r),
\label{eq:H_MF}
\end{equation}
where $H_0(\bm p)$ is the $10\times 10$ R5G Hamiltonian.  
Minimal coupling acts on each layer and sublattice component in the usual way.  
The mean-field problem is therefore that of a multicomponent Bloch electron in a uniform magnetic field and a moiré potential.

\section{Orbital content of the lowest Landau level}
\label{sec:LL_orbital_content}

The mean-field flux-attachment Hamiltonian acts on the full layer--sublattice spinor, so the resulting lowest conduction-band LL is not a single nonrelativistic LL orbital. Instead, it is a coherent superposition of different LL orbitals on different layer--sublattice components. For the chiral low-energy structure of R5G, the dominant components follow the sequence
\begin{equation}
|\Psi_{\rm LLL}\rangle
=
\sum_{\alpha=(l,\sigma)}
u_{\alpha}\, |n_{\alpha}\rangle_{\alpha},
\qquad
n_{\alpha}=(0,1,1,2,2,3,3,4,4,5)
\label{eq:supp_LLL_orbital_content}
\end{equation}
for $\alpha=(1A,1B,2A,2B,3A,3B,4A,4B,5A,5B)$, up to smaller corrections from trigonal warping and remote hopping. The weights $|u_\alpha|^2$ therefore show where the layer-resolved matrix elements have support in the LL wave function.

\begin{figure}[t]
    \centering
    \includegraphics[width=1.0\linewidth]{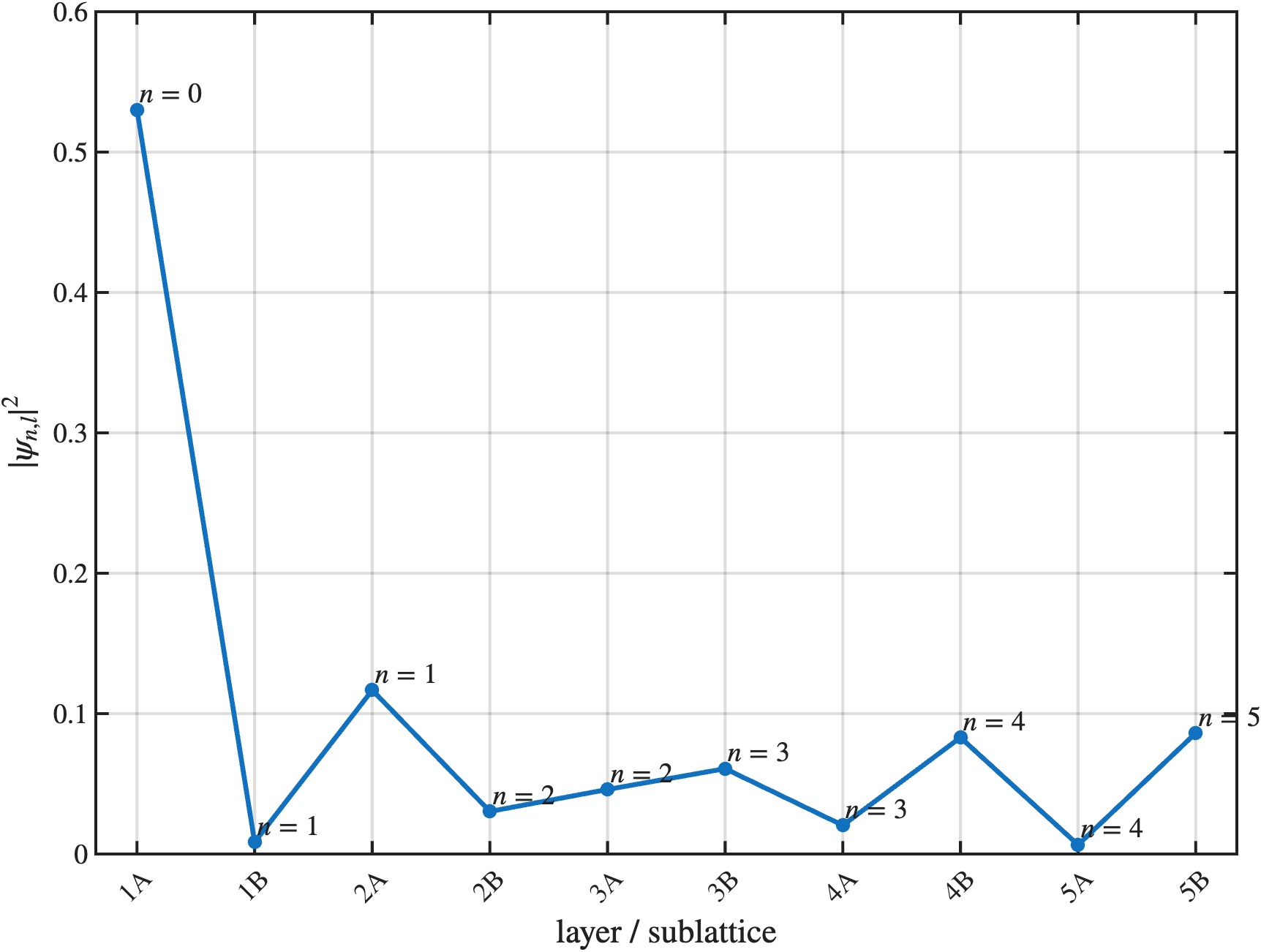}
    \caption{Selected layer--sublattice components of the lowest conduction-band LL wave function for $\Delta B=-\SI{24}{T}$ and $u_D=\SI{25}{meV}$. 
    The horizontal labels denote the layer--sublattice component, and the text labels indicate the LL orbital index $n_\alpha$ associated with that component. This density is shown as the orange trace in Fig.~3(b) of the main text. }
    \label{fig:supp-selected-LL-components}
\end{figure}

Figure~\ref{fig:supp-selected-LL-components} makes this structure explicit for $\Delta B=-\SI{24}{T}$ and $u_D=\SI{25}{meV}$. The largest component is the $1A$ orbital with $n=0$, confirming that the LLL remains anchored near the same layer--sublattice component as the zero-field conduction-band edge. 
The wave function also has smaller but visible weight on the remaining components, including $5A$ and $5B$ with higher LL orbital indices. 
The figure therefore shows directly where the probability weight of the R5G LL wave function resides.

\section{Periodic Landau-level basis and single-particle Hamiltonian}
\label{sec:pll_basis}

In this section we summarize the periodic Landau-level (PLL) basis used to represent the single-particle and interaction Hamiltonians in the presence of both a magnetic field and a moiré potential.

\subsection{Guiding-center and cyclotron coordinates}

For electrons in the perpendicular attached magnetic field $\Delta B\hat z$, the position operator can be decomposed as
\begin{equation}
\mathbf r = \mathbf R + \frac{\ell_B^2}{\hbar}\hat z \times \boldsymbol{\Pi},
\end{equation}
where $\boldsymbol{\Pi}=\mathbf p+e\mathbf A^*$ is the kinetic momentum and
$\ell_B=\sqrt{\hbar/(e|\Delta B|)}$ is the magnetic length. The sign of $\Delta B$ is fixed as in the main text; in the calculations below $\Delta B<0$.

The guiding-center coordinate $\mathbf R$ and cyclotron momentum $\boldsymbol{\Pi}$ satisfy
\begin{align}
[\Pi_x,\Pi_y] &= i\hbar^2/\ell_B^2,\\
[R_x,R_y] &= -i\ell_B^2,\\
[R_i,\Pi_j] &= 0 .
\end{align}
For this field direction we choose the oscillator phase convention
\begin{equation*}
a_{\rm LL}=-\frac{\ell_B}{\sqrt{2}\hbar}\Pi_+,
\qquad
a_{\rm LL}^\dagger=-\frac{\ell_B}{\sqrt{2}\hbar}\Pi_-,
\qquad [a_{\rm LL},a_{\rm LL}^\dagger]=1.
\end{equation*}
The minus signs are a basis convention, equivalent to $|n\rangle\to(-1)^n|n\rangle$ relative to the convention without them. They fix the phases of all Landau-orbital matrix elements below and are the convention used in the numerical calculation.

\subsection{Periodic Landau-level basis}

We consider a moiré lattice with primitive vectors $\mathbf a_{1,2}$ and reciprocal vectors $\mathbf G_{1,2}$ satisfying
$
\mathbf G_i\cdot \mathbf a_j = 2\pi\delta_{ij}.
$
We denote the magnitude of the magnetic flux per moiré unit cell by $p/q=|\Delta B|A_m/\Phi_0$, with $p,q>0$ coprime; the signed flux in our calculation is therefore $-p/q$. The magnetic translation operators are
\begin{equation}
T_{\mathbf a}=e^{i(\mathbf a\times\hat z)\cdot\mathbf R/\ell_B^2}.
\end{equation}

\begin{figure}[t]
    \centering
    \includegraphics[width=1.0\linewidth]{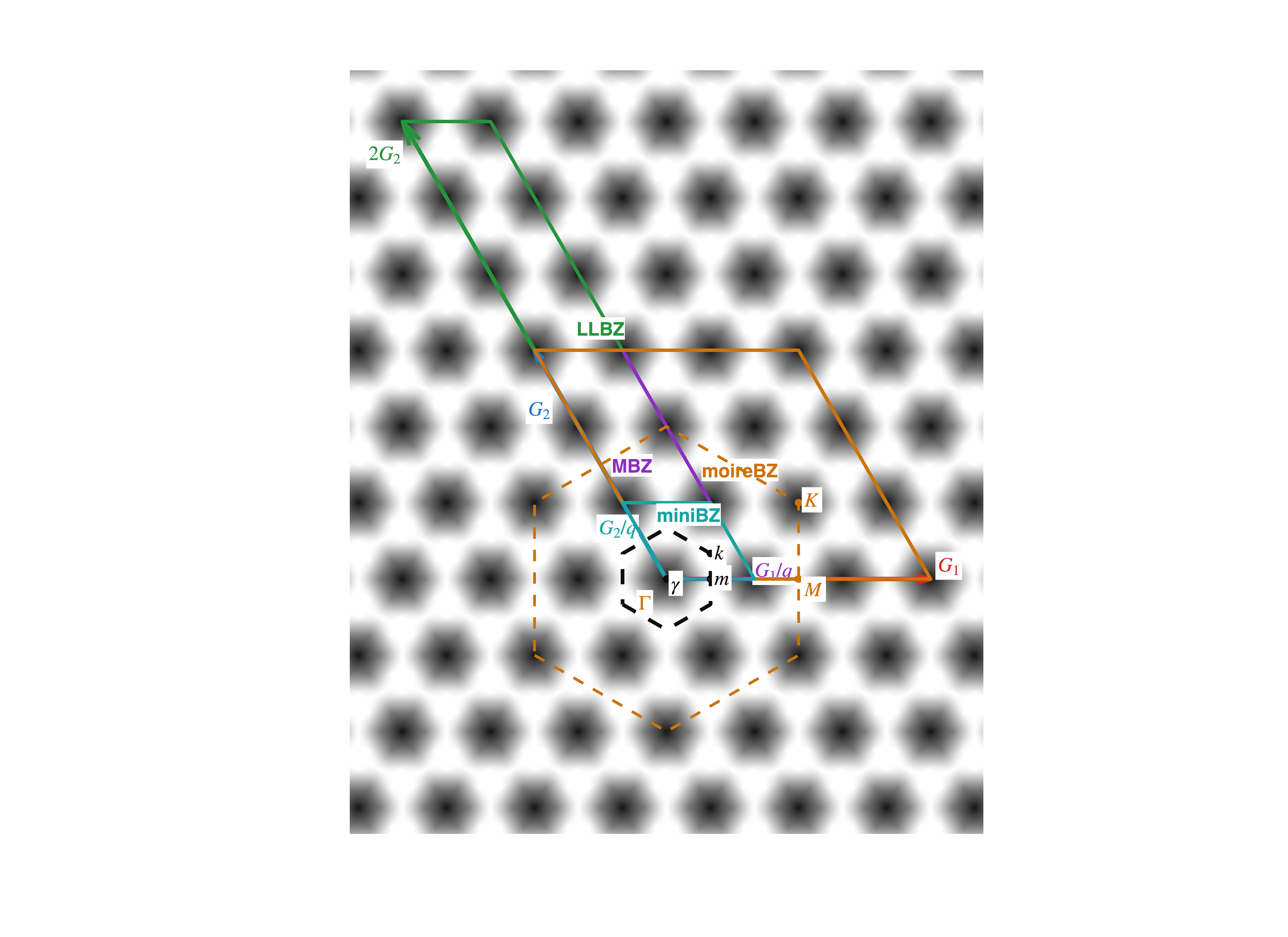}
    \caption{Reciprocal-space geometry and multiple Brillouin zones (BZs) for the representative case $\nu = 1/3$ ($p=2$, $q=3$). Colors encode the energy of the lowest HF band across multiple reciprocal-space copies. The black miniBZ Wigner--Seitz hexagon marks $\gamma$, $k$, and $m$; the orange electron moir\'e BZ hexagon marks $\Gamma$, $K$, and $M$. The parallelograms indicate the moir\'e BZ, MBZ, LLBZ, and miniBZ.}
    \label{fig:supp-hf-iter1-mbz-geometry}
\end{figure}

The periodic Landau-level basis $|n,\mathbf k\rangle$ is defined as simultaneous eigenstates of $\Pi^2$, $T_{q\mathbf a_1}$, and $T_{\mathbf a_2/p}$,
\begin{align}
\Pi^2 |n,\mathbf k\rangle &= E_n |n,\mathbf k\rangle,\\
T_{q\mathbf a_1}|n,\mathbf k\rangle &= e^{iq\mathbf a_1\cdot\mathbf k}|n,\mathbf k\rangle,\\
T_{\mathbf a_2/p}|n,\mathbf k\rangle &= e^{i(\mathbf a_2/p)\cdot\mathbf k}|n,\mathbf k\rangle.
\end{align}
It follows that the PLL states $|n,\mathbf k\rangle$, $|n,\mathbf k + \mathbf{G}_1 / q\rangle$, and $|n,\mathbf k + p\mathbf G_2\rangle$ are equivalent up to phase factors.
We therefore define the Landau-level Brillouin zone (LLBZ) as the parallelogram spanned by $\mathbf G_1/q$ and $p\mathbf G_2$, as shown in Fig.~\ref{fig:supp-hf-iter1-mbz-geometry} for $p=2$ and $q=3$.

The LLBZ momentum can be parameterized as
\begin{equation}
\mathbf k = k_1 \frac{1}{q}\mathbf G_1 + k_2\, p\mathbf G_2,
\qquad
k_1,k_2\in[0,1).
\end{equation}
A convenient explicit construction of the basis is \cite{kolavr2024hofstadter}
\begin{equation}
|n,\mathbf k\rangle =
e^{i k_2p\mathbf G_2\cdot\mathbf R}
e^{ik_1\frac1q\mathbf G_1\cdot\mathbf R}
|n,0\rangle .
\end{equation}
This expression applies for any ${\bf k}$ in the LLBZ. States with ${\bf k}$ outside the LLBZ are defined by periodic extension:
$|n,\mathbf k\rangle=|n,\mathbf k + \mathbf{G}_1 / q\rangle =|n,\mathbf k + p\mathbf G_2\rangle $.
This convention makes the basis discontinuous at the LLBZ boundary, but it is convenient for numerical implementation.
A continuous basis choice is also possible, as used in \cite{zou2025valley}.

For later calculations, we need the form factor $\langle n'\mathbf k'\ell'|e^{i\mathbf Q\cdot\mathbf r}|n\mathbf k\ell\rangle$.
Because the guiding-center and cyclotron operators commute, the Hilbert space factorizes into Landau-level and guiding-center sectors.
\begin{equation}\label{eq:form_factor}
\langle n'\mathbf k'\ell'|e^{i\mathbf Q\cdot\mathbf r}|n\mathbf k\ell\rangle
=
F_{n'n}(\mathbf Q)\,
\Lambda(\mathbf k',\mathbf k, \mathbf Q),
\end{equation}
with
\begin{equation}
F_{n'n}(\mathbf Q)
\equiv
\left\langle n'\left|\exp\!\left[-\frac{i\ell_B^2}{\hbar}(\hat z\times\mathbf Q)\cdot\boldsymbol{\Pi}\right]\right|n\right\rangle, \quad
\Lambda(\mathbf k',\mathbf k, \mathbf Q) = \left\langle \mathbf{k}^{\prime}\right| e^{i \mathbf{Q} \cdot \mathbf{R}}|\mathbf{k}\rangle.
\end{equation}
Using the oscillator convention above, the cyclotron form factor is \cite{murthy2003hamiltonian}
\begin{equation}
F_{n'n}(\mathbf Q)
=
\begin{cases}
e^{-Q^2\ell_B^2/4}
\sqrt{\dfrac{n!}{n'!}}
\left(-\dfrac{\ell_B(Q_x+iQ_y)}{\sqrt2}\right)^{n'-n}
L_{n}^{\,n'-n}\!\left(\dfrac{Q^2\ell_B^2}{2}\right), & n'\ge n,\\[8pt]
\big[F_{nn'}(-\mathbf Q)\big]^*, & n'<n .
\end{cases}
\end{equation}
The guiding-center form factor follows from the definition of the PLL basis and the algebra of magnetic translations:
\begin{equation}
\left\langle \mathbf{k}^{\prime}\right| e^{i \mathbf{Q} \cdot \mathbf{R}}|\mathbf{k}\rangle
=\delta_{\mathbf{k}^{\prime},\, \mathbf{k}+\mathbf{Q}}^{(\mathrm{LLBZ})}
\exp \left[-i \pi Q_1 Q_2-i 2 \pi k_1\left\lfloor k_2+Q_2\right\rfloor+i 2 \pi Q_1\left\{k_2+Q_2\right\}\right].
\end{equation}
Here $\delta_{\mathbf{k}^{\prime},\, \mathbf{k}+\mathbf{Q}}^{(\mathrm{LLBZ})}$ means that $\mathbf{k}'$ and $\mathbf{k}+\mathbf{Q}$ are equivalent up to LLBZ reciprocal lattice vectors.

\subsection{Single-particle moir\'e Hamiltonian}

The moiré potential
\begin{equation}
V(\mathbf r)=\sum_{\mathbf g} V_{\mathbf g} e^{i\mathbf g\cdot\mathbf r},
\qquad
\mathbf g=m_1\mathbf G_1+m_2\mathbf G_2,
\end{equation}
couples the PLL states $|n,\mathbf k\rangle$ and $|n',\mathbf k+\mathbf g\rangle$.
This coupling folds the LLBZ into the magnetic Brillouin zone (MBZ), spanned by $\mathbf G_1/q$ and $\mathbf G_2$, as shown in Fig.~\ref{fig:supp-hf-iter1-mbz-geometry}, and produces $p$ LL orbitals for each ${\bf \kappa}$ in the MBZ.
We therefore label the LLBZ momentum $\mathbf k$ as ${\bf \kappa} + \ell \mathbf G_2$, with ${\bf \kappa}$ in the MBZ and $\ell=0,\dots,p-1$.

Including internal orbital indices $\alpha$ (layer, sublattice, etc.), the field operator is expanded as
\begin{equation}
\psi_\alpha(\mathbf r)
=
\sum_{n,{\bf \kappa},\ell}
\phi_{n{\bf \kappa}\ell}(\mathbf r)
\, c_{n{\bf \kappa}\ell\alpha},
\end{equation}
where $c_{n{\bf \kappa}\ell\alpha}$ annihilates the folded PLL basis state at MBZ momentum ${\bf \kappa}$.
The single-particle Hamiltonian consists of the Landau-level kinetic energy and the moiré potential,
\begin{equation}
H_{\mathrm{sp}}=H_{\mathrm{kin}}+H_{\mathrm{moire}} .
\end{equation}
For R5G, the kinetic term is a matrix in Landau-orbital and layer--sublattice indices $(n,\alpha)$ but is independent of the guiding-center momentum and diagonal in the folded index $\ell$:
\begin{equation}
H_{\mathrm{kin}}
=
\sum_{{\bf \kappa}}
\sum_{n'\ell'\alpha'}\sum_{n\ell\alpha}
\big[H_{\mathrm{R5G}}^{(B)}\big]_{n'\alpha',\,n\alpha}\,
\delta_{\ell'\ell}\,
c^\dagger_{n'{\bf \kappa}\ell'\alpha'}
c_{n{\bf \kappa}\ell\alpha}.
\end{equation}
The moiré potential gives the matrix elements
\begin{equation}
H_{\mathrm{moire}}
=
\sum_{\mathbf g}\sum_{{\bf \kappa}}
\sum_{\ell\ell'}\sum_{n'n}\sum_{\alpha\beta}
V_{\mathbf g}
[P_{l_0}]_{\alpha\beta}
\langle n'{\bf \kappa}\ell'|e^{i\mathbf g\cdot\mathbf r}|n{\bf \kappa}\ell\rangle
c^\dagger_{n'{\bf \kappa}\ell'\alpha}
c_{n{\bf \kappa}\ell\beta}.
\end{equation}
Here
$[P_{l_0}]_{\alpha\beta}=\delta_{\alpha\beta}\delta_{l(\alpha),l_0}$
projects onto the two sublattices of the graphene layer adjacent to hBN. Thus the moiré potential is diagonal in the internal orbital index and vanishes on all other layers.
For this matrix element, we use Eq.~\eqref{eq:form_factor} with $\mathbf Q = \mathbf g$.
For the folded $({\bf \kappa},\ell)$ basis used below, ${\bf k}\rightarrow{\bf \kappa} + \ell\mathbf G_2$, the guiding-center part becomes
\begin{equation}
\Lambda_{\ell'\ell}({\bf \kappa}',{\bf \kappa}, \mathbf Q)
=
\delta_{{\bf \kappa}' + \ell'\mathbf{G}_2,\,{\bf \kappa} + \ell \mathbf{G}_2+\mathbf{Q}}^{(\mathrm{LLBZ})}
\,e^{i\Phi_{\mathbf Q}({\bf \kappa},\ell)},
\end{equation}
The $\delta$ function reduces to $\delta_{{\bf \kappa}' + \ell'\mathbf{G}_2,\,{\bf \kappa} + \ell \mathbf{G}_2+\mathbf{Q}}^{(\mathrm{LLBZ})} = \sum_{\mathbf{G} \in \textrm{mBZ}} \delta_{\mathbf{Q}, {\bf \kappa}' - {\bf \kappa} - \mathbf{G}} \delta_{\ell', \ell - m(G)}$,
where $m(\mathbf{G})$ is the integer part of the $G_2$ component of $\mathbf G$, and
\begin{equation}
\Phi_{\mathbf Q}({\bf \kappa},\ell)
=
-\pi Q_1Q_2
-2\pi \kappa_1 \lfloor \kappa_2 + \ell / p +Q_2\rfloor
+2\pi Q_1\{\kappa_2 + \ell / p +Q_2\}.
\end{equation}
Here ${\bf \kappa} = \kappa_1\mathbf G_1 / q + \kappa_2p\mathbf G_2$.

Restricting to moiré reciprocal vectors $\mathbf g = m_1\mathbf G_1 + m_2\mathbf G_2$ gives
\begin{equation}
\Lambda_{\ell'\ell}({\bf \kappa},\mathbf g)
=
\delta_{\ell',\,\ell+m_2 \,(\mathrm{mod}\,p)}
\,e^{i\Phi_{\mathbf g}({\bf \kappa},\ell)},
\end{equation}
with
\begin{equation}
\Phi_{\mathbf g}({\bf \kappa},\ell)
=
-\pi m_1m_2q / p
-2\pi \kappa_1\lfloor \kappa_2 + \ell / p+m_2 / p\rfloor
+2\pi m_1 q\{\kappa_2+ \ell / p + m_2 /p\}.
\end{equation}
This reproduces the expressions of Ref.~\cite{kolavr2024hofstadter}.

\subsection{Chern number from a small Wilson loop}\label{sec:Berry}

We consider a normalized isolated band state
\begin{equation}
|\Psi_\lambda({\bf \kappa})\rangle
=
\sum_{a}
z^{(\lambda)}_{a}({\bf \kappa})\,
|a,{\bf \kappa}\rangle ,
\end{equation}
where $a$ is a collective index for the PLL basis at fixed folded-MBZ momentum ${\bf \kappa}$; for example, it may denote the tuple $(n,\ell,\alpha)$ appearing above. Because the basis itself depends on momentum, the Berry connection is not obtained from the naive coefficient overlap $z_\lambda^\dagger({\bf \kappa}+\delta{\bf \kappa})\,z_\lambda({\bf \kappa})$. Instead, one must use the physical overlap between neighboring states, which is naturally expressed through the density-matrix element of $e^{i\delta\mathbf q\cdot\mathbf r}$.

For a small step $\delta{\bf \kappa}$, let $\delta\mathbf q$ be the corresponding physical momentum transfer. The link variable is defined by
\begin{equation}
\mathcal U_\lambda({\bf \kappa},\delta{\bf \kappa})
\equiv
\langle \Psi_\lambda({\bf \kappa}+\delta{\bf \kappa})|\,e^{i\delta\mathbf q\cdot\mathbf r}\,|\Psi_\lambda({\bf \kappa})\rangle .
\end{equation}
Using the form factor
\begin{equation}
M_{ab}({\bf \kappa},\delta\mathbf q)
=
\langle a,{\bf \kappa}+\delta{\bf \kappa}|
\,e^{i\delta\mathbf q\cdot\mathbf r}\,
|b,{\bf \kappa}\rangle ,
\end{equation}
the corresponding $U(1)$ link is
\begin{equation}
U_\lambda({\bf \kappa},\delta{\bf \kappa})
=
\frac{
z_\lambda^\dagger({\bf \kappa}+\delta{\bf \kappa})\,
M({\bf \kappa},\delta\mathbf q)\,
z_\lambda({\bf \kappa})}
{\big|z_\lambda^\dagger({\bf \kappa}+\delta{\bf \kappa})\,
M({\bf \kappa},\delta\mathbf q)\,
z_\lambda({\bf \kappa})\big|}.
\end{equation}
This link is ordered as $\langle\Psi({\bf \kappa}+\delta{\bf \kappa})|\cdots|\Psi({\bf \kappa})\rangle$, the complex conjugate of the forward-overlap ordering often used in the Fukui--Hatsugai convention. Consequently, the sign of $C$ defined below is the opposite of that convention. We use the present convention consistently throughout: a filled parent LLL at $\Delta B<0$ has $C=-1$, and the electronic Hall response is $\sigma_{xy}=-[C/(2sC+1)]e^2/h$.

For a plaquette spanned by two elementary steps $\delta{\bf \kappa}_1$ and $\delta{\bf \kappa}_2$, the small Wilson loop is
\begin{equation}
W_\lambda({\bf \kappa})
=
U_\lambda({\bf \kappa},\delta{\bf \kappa}_1)\,
U_\lambda({\bf \kappa}+\delta{\bf \kappa}_1,\delta{\bf \kappa}_2)\,
U_\lambda({\bf \kappa}+\delta{\bf \kappa}_2,\delta{\bf \kappa}_1)^{-1}\,
U_\lambda({\bf \kappa},\delta{\bf \kappa}_2)^{-1}.
\end{equation}
The discrete Berry curvature on the plaquette is
\begin{equation}
\mathcal F_\lambda({\bf \kappa})=\arg W_\lambda({\bf \kappa}),
\end{equation}
and the Chern number is
\begin{equation}
C_\lambda=\frac{1}{2\pi}\sum_{{\bf \kappa}}\mathcal F_\lambda({\bf \kappa}),
\end{equation}
where the sum runs over all plaquettes of the folded MBZ. Equivalently, translation covariance divides the folded MBZ into $q$ symmetry-related copies of the translational miniBZ, so one may integrate over a single miniBZ Wigner--Seitz cell and multiply the result by $q$. This is the procedure used for the numerical Berry-curvature plots.

This formulation is gauge invariant and automatically includes the momentum dependence and quasiperiodicity of the basis through $M({\bf \kappa},\delta\mathbf q)$. In particular, it reduces to the usual Landau-level result in the trivial-band limit. Indeed, if the band state coincides with a single Landau-level basis state, then for sufficiently small $\delta\mathbf q$ (so that no fold-back occurs in the LLBZ),
\begin{equation}
M({\bf \kappa},\delta\mathbf q)
\simeq
F_n(\delta\mathbf q)\,
\exp\!\left[
i2\pi\,\delta q_1 \kappa_2
+i\pi\,\delta q_1\delta q_2
\right],
\end{equation}
where $\delta\mathbf q=\delta q_1\,\mathbf G_1/q+\delta q_2\,p\mathbf G_2$, and $F_n(\delta\mathbf q)$ is real to leading order. The usual Landau-level Berry phase therefore comes entirely from the guiding-center part of $M$, and the plaquette Wilson loop is proportional to the enclosed momentum-space area, giving the familiar uniform Berry curvature of a Landau level. Thus $M({\bf \kappa},\delta\mathbf q)$ is the natural generalization of the ordinary Landau-level overlap to a general projected band.

\subsection{Symmetries of the moir\'e system}

Although the folded Hamiltonian is defined on the MBZ spanned by $\mathbf G_1/q$ and $\mathbf G_2$, this MBZ is not the smallest periodic cell of the spectrum.  The reason is that the primitive moir\'e translation along $\mathbf a_1$ remains a symmetry.  In the periodic Landau-level basis, this translation shifts the momentum by $-p\mathbf G_2/q$ (folded back into the MBZ) and acts on the folded orbital index through a unitary matrix $U_{a_1}({\bf \kappa})$.  Consequently, the Hamiltonian is not strictly periodic as a matrix under this shift, but is covariant:
\begin{equation}
H_{\mathrm{sp}}({\bf \kappa}')
=
U_{\bf a_1}({\bf \kappa})\,
H_{\mathrm{sp}}({\bf \kappa})\,
U_{\bf a_1}^\dagger({\bf \kappa}) .
\end{equation}
Since $p$ and $q$ are coprime, repeated translations generate the same set of MBZ copies as shifts by $\mathbf G_2/q$.  This unitary covariance makes the spectrum periodic on the miniBZ spanned by $\mathbf G_1/q$ and $\mathbf G_2/q$, as illustrated in Fig.~\ref{fig:supp-hf-iter1-mbz-geometry}.

The folded basis also preserves the threefold rotation symmetry.  Under $C_3$, a basis state at ${\bf \kappa}$ is mapped to the folded representative of $C_3{\bf \kappa}$, again with a momentum-dependent permutation and phase acting within the folded orbital space.  For a $C_3$-symmetric moir\'e Hamiltonian,
\begin{equation}
H_{\mathrm{sp}}({\bf \kappa}_{C_3})
=
U_{C_3}({\bf \kappa})\,
H_{\mathrm{sp}}({\bf \kappa})\,
U_{C_3}^\dagger({\bf \kappa}),
\end{equation}
where ${\bf \kappa}_{C_3}$ denotes the rotated momentum folded back to the MBZ.  The apparent reshuffling of components is therefore only a basis effect: the Hamiltonian and its spectrum preserve the moir\'e translation and $C_3$ symmetries.

\section{Projected density operator and Hartree--Fock formalism}

\subsection{Density operator and interaction Hamiltonian}

The Fourier component of the density operator is
\begin{equation}
\rho_{\mathbf Q}
=
\sum_\alpha
\int d^2r\, e^{i\mathbf Q\cdot\mathbf r}
\psi_\alpha^\dagger(\mathbf r)\psi_\alpha(\mathbf r).
\end{equation}

Substituting the field-operator expansion gives the second-quantized form
\begin{equation}
\rho_{\mathbf Q}
=
\sum_{{\bf \kappa}',{\bf \kappa}}\sum_{n'n}\sum_{\ell'\ell}\sum_\alpha
\langle n'{\bf \kappa}'\ell'|e^{i\mathbf Q\cdot\mathbf r}|n{\bf \kappa}\ell\rangle\,
c^\dagger_{n'{\bf \kappa}'\ell'\alpha}
c_{n{\bf \kappa}\ell\alpha}.
\end{equation}


The Coulomb interaction is written as
\begin{equation}
H_{\mathrm{int}}
=
\frac{1}{2A}
\sum_{\mathbf Q}
V(\mathbf Q)
:\!\rho_{-\mathbf Q}\rho_{\mathbf Q}\!:.
\end{equation}
For a two-dimensional Coulomb interaction in momentum space, we use
$
V(\mathbf Q)
=
\frac{2\pi e^2}{\epsilon |\mathbf Q|},
$
where $\epsilon$ is the dielectric constant.
A uniform neutralizing background removes the spatially uniform Hartree component; in the finite-system sums we implement this by setting $V(\mathbf Q=0)=0$. The normal ordering removes the one-body self-contraction. The remaining integrable small-$Q$ exchange singularity is treated with the same finite-size regularization in all calculations.

Substituting the density operator gives
\begin{align}
H_{\mathrm{int}}
=
\frac{1}{2A}
\sum_{\mathbf Q}
V(\mathbf Q)
\sum_{{\bf \kappa}_1,{\bf \kappa}_2,{\bf \kappa}_3,{\bf \kappa}_4}
\sum_{\ell_i}
\sum_{n_i}
\sum_{\alpha\beta}
\langle n_1{\bf \kappa}_1\ell_1|e^{-i\mathbf Q\cdot\mathbf r}|n_4{\bf \kappa}_4\ell_4\rangle
\langle n_2{\bf \kappa}_2\ell_2|e^{i\mathbf Q\cdot\mathbf r}|n_3{\bf \kappa}_3\ell_3\rangle
\nonumber\\
\times
c^\dagger_{n_1{\bf \kappa}_1\ell_1\alpha}
c^\dagger_{n_2{\bf \kappa}_2\ell_2\beta}
c_{n_3{\bf \kappa}_3\ell_3\beta}
c_{n_4{\bf \kappa}_4\ell_4\alpha}.
\end{align}

The single-particle Hilbert space at each ${\bf \kappa}$ is finite after truncating the Landau-level index $n$ and is labeled additionally by the guiding-center orbital $\ell=0,\dots,p-1$ and internal orbital index $\alpha$; the interaction transfers momentum between these ${\bf \kappa}$ sectors.

\subsection{Projection to a subset of kinetic bands}

Diagonalize the uniform-field kinetic matrix in $(n,\alpha)$:
\begin{equation}
\sum_{n\alpha}
\big[h_{\mathrm{kin}}^{(B)}\big]_{n'\alpha',\,n\alpha}\,
U_{n\alpha}^{\,a}
=
\varepsilon_{a}\,
U_{n'\alpha'}^{\,a},
\end{equation}
which is independent of ${\bf \kappa}$ and $\ell$. Define band operators
\begin{equation}
d_{a{\bf \kappa}\ell}
=
\sum_{n\alpha}
\big[U_{n\alpha}^{\,a}\big]^*\,
c_{n{\bf \kappa}\ell\alpha}.
\end{equation}
Using the factorized matrix element
$\langle n'{\bf \kappa}\ell'|e^{i\mathbf Q\cdot\mathbf r}|n{\bf \kappa}\ell\rangle
=F_{n'n}(\mathbf Q)\Lambda_{\ell'\ell}({\bf \kappa},\mathbf Q)$ from above, the density operator in this basis is
\begin{equation}
\rho_{\mathbf Q}
=
\sum_{{\bf \kappa}}\sum_{\ell\ell'}\sum_{ab}
\mathcal M_{a\ell',\,b\ell}({\bf \kappa},\mathbf Q)\,
d_{a,{\bf \kappa}+\mathbf Q,\ell'}^\dagger
d_{b,{\bf \kappa},\ell},
\end{equation}
with
\begin{equation}
\mathcal M_{a\ell',\,b\ell}({\bf \kappa},\mathbf Q)
=
\sum_{\substack{n'n\\\alpha}}
\big[U_{n'\alpha}^{\,a}\big]^*
F_{n'n}(\mathbf Q)\,
\Lambda_{\ell'\ell}({\bf \kappa},\mathbf Q)\,
U_{n\alpha}^{\,b}.
\end{equation}

We then choose a subset $\mathcal S$ of kinetic bands, for example the lowest few bands, and project:
\begin{equation}
\bar{\rho}_{\mathbf Q}
=
\sum_{{\bf \kappa}}\sum_{\ell\ell'}\sum_{a,b\in\mathcal S}
\mathcal M_{a\ell',\,b\ell}({\bf \kappa},\mathbf Q)\,
d_{a,{\bf \kappa}+\mathbf Q,\ell'}^\dagger
d_{b,{\bf \kappa},\ell}.
\end{equation}
The corresponding projected interaction is
\begin{equation}
H_{\mathrm{int}}^{(\mathcal S)}
=
\frac{1}{2A}\sum_{\mathbf Q}
V(\mathbf Q)\,
:\!\bar{\rho}_{-\mathbf Q}\bar{\rho}_{\mathbf Q}\!:,
\end{equation}
or explicitly
\begin{align}
H_{\mathrm{int}}^{(\mathcal S)}
=
\frac{1}{2A}
\sum_{\mathbf Q,{\bf \kappa},{\bf \kappa}'}
\sum_{\ell_1\ell_2\ell_3\ell_4}
\sum_{\substack{a,b\in\mathcal S\\c,d\in\mathcal S}}
V(\mathbf Q)\,
\mathcal M_{a\ell_1,\,b\ell_2}({\bf \kappa},\mathbf Q)\,
\mathcal M_{c\ell_3,\,d\ell_4}({\bf \kappa}',-\mathbf Q)
\nonumber\\
\times
d^\dagger_{a,{\bf \kappa}+\mathbf Q,\ell_1}
d^\dagger_{c,{\bf \kappa}'-\mathbf Q,\ell_3}
d_{d,{\bf \kappa}',\ell_4}
d_{b,{\bf \kappa},\ell_2}.
\end{align}

\subsection{Hartree--Fock decoupling in the projected single-LL basis}
\label{sec:HF_projected_singleLL}

We now restrict to a single projected Landau level, or more generally to a single flat kinetic band, and suppress the band label $a_0$.  The remaining Hilbert space at fixed ${\bf \kappa}\in \mathrm{MBZ}$ is spanned by the folded orbital index $\ell=0,\dots,p-1$, with annihilation operators $d_{\ell,{\bf \kappa}}$.

Because the underlying Landau level is flat, the projected spinor $U_{n\alpha}$ is independent of both ${\bf \kappa}$ and $\ell$.  The projected density operator can therefore be written as
\begin{equation}
\bar\rho_{\bm Q}
=
\sum_{{\bf \kappa}\in \mathrm{MBZ}}
\sum_{\ell_1,\ell_2=0}^{p-1}
M_{\ell_1\ell_2}({\bf \kappa},\bm Q)\,
d^\dagger_{\ell_1,{\bf \kappa}+\bm Q}\,
d_{\ell_2,{\bf \kappa}},
\label{eq:rho_projected_singleLL}
\end{equation}
with projected form factor
\begin{equation}
M_{\ell_1\ell_2}({\bf \kappa},\bm Q)
=
\mathcal F(\bm Q)\,
\Lambda_{\ell_1\ell_2}({\bf \kappa},\bm Q),
\label{eq:M_singleLL}
\end{equation}
where
\begin{equation}
\mathcal F(\bm Q)
=
\sum_{n'n,\alpha}
U^{*}_{n'\alpha}\,
F_{n'n}(\bm Q)\,
U_{n\alpha}.
\label{eq:Fproj_singleLL}
\end{equation}
Here $F_{n'n}(\bm Q)$ is the cyclotron form factor and $\Lambda_{\ell_1\ell_2}({\bf \kappa},\bm Q)$ is the guiding-center form factor.

The projected interaction Hamiltonian is
\begin{equation}
H_{\mathrm{int}}^{\mathrm{proj}}
=
\frac{1}{2A}
\sum_{\bm Q}
V(\bm Q)\,
:\!\bar\rho_{-\bm Q}\bar\rho_{\bm Q}\!:.
\label{eq:Hint_projected_singleLL}
\end{equation}

\paragraph{Order parameter.}
In the folded basis the natural Hartree--Fock order parameter is the equal-momentum density matrix
\begin{equation}
\rho_{\ell\ell'}({\bf \kappa})
\equiv
\left\langle
d^\dagger_{\ell,{\bf \kappa}}\,
d_{\ell',{\bf \kappa}}
\right\rangle .
\label{eq:rho_order_parameter_singleLL}
\end{equation}
For a translationally invariant or MBZ-periodic Hartree--Fock state,
\begin{equation}
\left\langle
d^\dagger_{\ell,{\bf \kappa}+\bm Q}\,
d_{\ell',{\bf \kappa}}
\right\rangle
=
\delta_{\bm Q,\bm G}\,
\rho_{\ell+m(\bm G),\,\ell'}({\bf \kappa}),
\qquad
\bm G\in \mathcal G_{\mathrm{MBZ}},
\label{eq:selection_rule_singleLL}
\end{equation}
where
\begin{equation}
\bm G = m_1 \frac{\bm G_1}{q} + m_2 \bm G_2,
\qquad
m(\bm G)\equiv m_2 \pmod p,
\label{eq:mG_singleLL}
\end{equation}
and periodicity in the folded basis implies
\begin{equation}
d_{\ell,{\bf \kappa}+\bm G}=d_{\ell+m(\bm G),\,{\bf \kappa}}.
\label{eq:folded_periodicity_singleLL}
\end{equation}

\paragraph{Hartree term.}
Using Eq.~\eqref{eq:selection_rule_singleLL},
\begin{equation}
\langle \bar\rho_{-\bm G}\rangle
=
\sum_{{\bf \kappa}'\in \mathrm{MBZ}}
\sum_{a,b=0}^{p-1}
M_{a+m(\bm G),\,b}({\bf \kappa}',-\bm G)\,
\rho_{ab}({\bf \kappa}').
\label{eq:rhoG_expectation_singleLL}
\end{equation}
The Hartree contribution is then
\begin{equation}
H_{\mathrm H}
=
\frac{1}{A}
\sum_{\bm G\in \mathcal G_{\mathrm{MBZ}}}
V(\bm G)\,
\langle \bar\rho_{-\bm G}\rangle\,
\bar\rho_{\bm G}.
\label{eq:Hartree_start_singleLL}
\end{equation}
Here and below the $\bm G=0$ Hartree component is understood to vanish because of the neutralizing background specified above.
Reducing $d^\dagger_{\ell,{\bf \kappa}+\bm G}$ back to the MBZ using Eq.~\eqref{eq:folded_periodicity_singleLL}, we obtain
\begin{equation}
H_{\mathrm H}
=
\sum_{{\bf \kappa}\in \mathrm{MBZ}}
\sum_{\ell,\ell'=0}^{p-1}
\big[h_{\mathrm H}({\bf \kappa})\big]_{\ell\ell'}\,
d^\dagger_{\ell,{\bf \kappa}}\,
d_{\ell',{\bf \kappa}},
\label{eq:Hartree_matrix_form_singleLL}
\end{equation}
with
\begin{equation}
\big[h_{\mathrm H}({\bf \kappa})\big]_{\ell\ell'}
=
\sum_{\bm G\in \mathcal G_{\mathrm{MBZ}}}
\sum_{{\bf \kappa}'\in \mathrm{MBZ}}
\sum_{a,b=0}^{p-1}
K^{\mathrm H}_{ab;\,\ell\ell'}({\bf \kappa},{\bf \kappa}';\bm G)\,
\rho_{ab}({\bf \kappa}'),
\label{eq:Hartree_kernel_contraction_singleLL}
\end{equation}
where the Hartree kernel is
\begin{equation}
K^{\mathrm H}_{ab;\,\ell\ell'}({\bf \kappa},{\bf \kappa}';\bm G)
=
\frac{1}{A}\,
V(\bm G)\,
M_{\ell-m(\bm G),\,\ell'}({\bf \kappa},\bm G)\,
M_{a+m(\bm G),\,b}({\bf \kappa}',-\bm G).
\label{eq:Hartree_kernel_singleLL}
\end{equation}

\paragraph{Fock term.}
Starting from Eq.~\eqref{eq:Hint_projected_singleLL}, the exchange contraction of
$d^\dagger_{\ell_1,{\bf \kappa}-\bm Q}$ with $d_{\ell_4,{\bf \kappa}'}$ is nonzero only if
\begin{equation}
{\bf \kappa}-\bm Q = {\bf \kappa}' - \bm G,
\qquad
\bm G\in \mathcal G_{\mathrm{MBZ}},
\label{eq:Fock_constraint_singleLL}
\end{equation}
or equivalently
\begin{equation}
\bm q^{\,\bm G}({\bf \kappa},{\bf \kappa}')
\equiv
{\bf \kappa}' - {\bf \kappa} - \bm G,
\qquad
\bm Q = - \bm q^{\,\bm G}({\bf \kappa},{\bf \kappa}').
\label{eq:qG_definition_singleLL}
\end{equation}
Using Eq.~\eqref{eq:folded_periodicity_singleLL},
\begin{equation}
\left\langle
d^\dagger_{\ell_1,{\bf \kappa}-\bm Q}\,
d_{\ell_4,{\bf \kappa}'}
\right\rangle
=
\rho_{\ell_1-m(\bm G),\,\ell_4}({\bf \kappa}').
\label{eq:Fock_expectation_singleLL}
\end{equation}
The Fock contribution can therefore be written as
\begin{equation}
H_{\mathrm F}
=
\sum_{{\bf \kappa}\in \mathrm{MBZ}}
\sum_{\ell,\ell'=0}^{p-1}
\big[h_{\mathrm F}({\bf \kappa})\big]_{\ell\ell'}\,
d^\dagger_{\ell,{\bf \kappa}}\,
d_{\ell',{\bf \kappa}},
\label{eq:Fock_matrix_form_singleLL}
\end{equation}
with
\begin{equation}
\big[h_{\mathrm F}({\bf \kappa})\big]_{\ell\ell'}
=
\sum_{\bm G\in \mathcal G_{\mathrm{MBZ}}}
\sum_{{\bf \kappa}'\in \mathrm{MBZ}}
\sum_{a,b=0}^{p-1}
K^{\mathrm F}_{ab;\,\ell\ell'}({\bf \kappa},{\bf \kappa}';\bm G)\,
\rho_{ab}({\bf \kappa}'),
\label{eq:Fock_kernel_contraction_singleLL}
\end{equation}
where the exchange kernel is
\begin{equation}
K^{\mathrm F}_{ab;\,\ell\ell'}({\bf \kappa},{\bf \kappa}';\bm G)
=
-\frac{1}{A}\,
V\!\left(\bm q^{\,\bm G}({\bf \kappa},{\bf \kappa}')\right)\,
M_{a+m(\bm G),\,\ell'}\!\left({\bf \kappa},\bm q^{\,\bm G}({\bf \kappa},{\bf \kappa}')\right)\,
M_{\ell-m(\bm G),\,b}\!\left({\bf \kappa}',-\bm q^{\,\bm G}({\bf \kappa},{\bf \kappa}')\right).
\label{eq:Fock_kernel_singleLL}
\end{equation}

\paragraph{Total Hartree--Fock Hamiltonian.}
Combining the projected single-particle term $h_{\mathrm{sp}}({\bf \kappa})$ with the Hartree and Fock contributions gives
\begin{equation}
H_{\mathrm{HF}}
=
\sum_{{\bf \kappa}\in \mathrm{MBZ}}
\sum_{\ell,\ell'=0}^{p-1}
\Big(
[h_{\mathrm{sp}}({\bf \kappa})]_{\ell\ell'}
+
[h_{\mathrm H}({\bf \kappa})]_{\ell\ell'}
+
[h_{\mathrm F}({\bf \kappa})]_{\ell\ell'}
\Big)
d^\dagger_{\ell,{\bf \kappa}}d_{\ell',{\bf \kappa}}.
\label{eq:HF_total_singleLL_corrected}
\end{equation}
At each ${\bf \kappa}$, $H_{\mathrm{HF}}({\bf \kappa})$ is a $p\times p$ Hermitian matrix in the folded-orbital space. Diagonalizing Eq.~\eqref{eq:HF_total_singleLL_corrected} yields the Hartree--Fock quasiparticle bands and eigenvectors, from which the density matrix \eqref{eq:rho_order_parameter_singleLL} is updated self-consistently.

\section{Hartree--Fock results}\label{sec:HF_results}

\begin{figure}[htbp]
    \centering
    \includegraphics[width=1.0\linewidth]{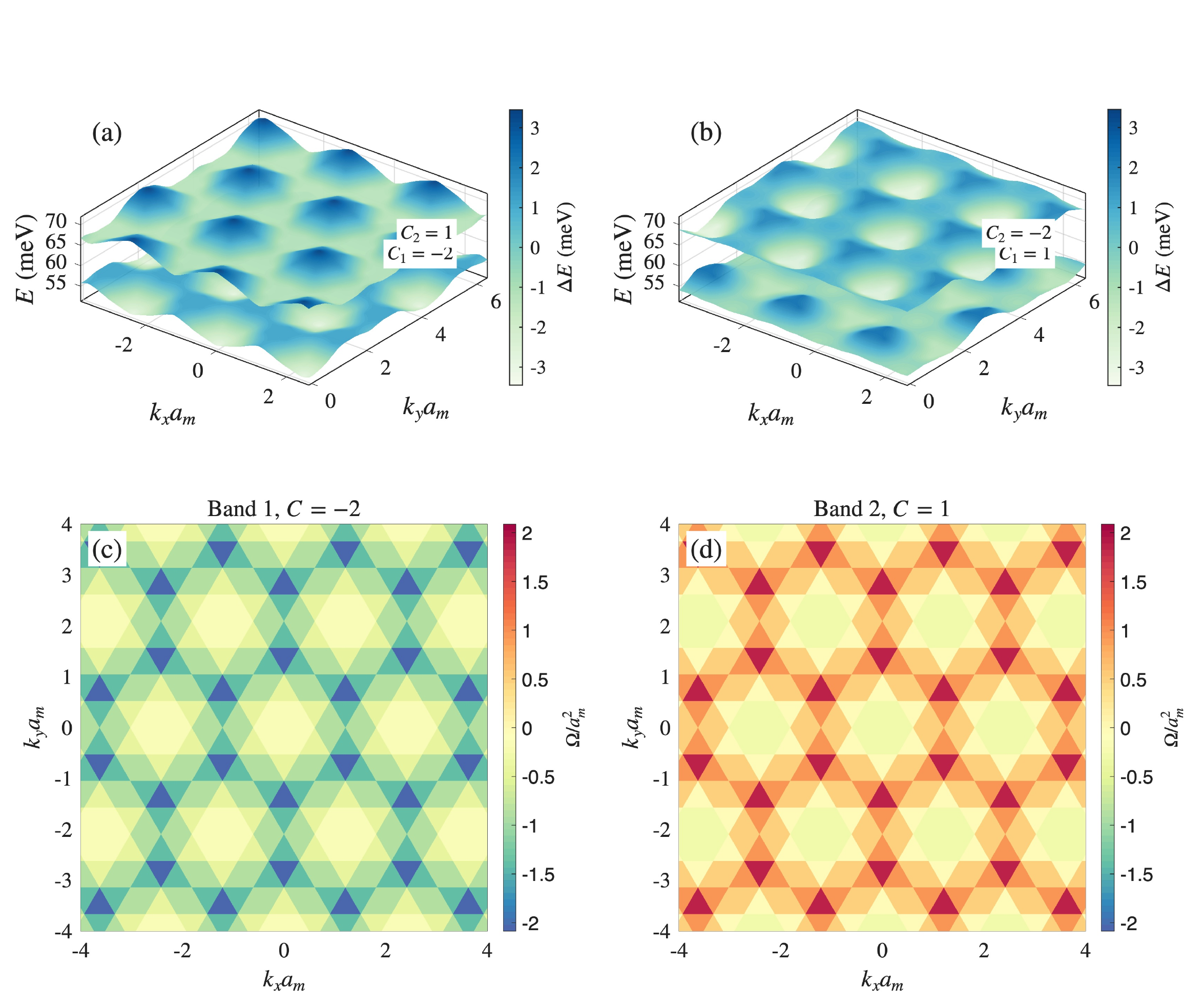}
    \caption{Converged Hartree--Fock band structures and Berry curvature at $\nu=1/3$, with physical moir\'e phase $\phi=0$, $V_m=\SI{20}{meV}$ on the remote layer, dielectric constant $\epsilon=5$, and the two lowest conduction LLs retained in the calculation. Panels (a) and (b) show the two lowest HF subbands obtained from $\phi_{\rm seed}=0$ and $\phi_{\rm seed}=\pi/3$, respectively. Their Chern numbers in ascending-energy order are $(-2,1)$ and $(1,-2)$. The color scale shows $\Delta E_n({\bf k})\equiv E_n({\bf k})-\overline{E}_n$, the energy of each band relative to its mean. Panels (c) and (d) show the Berry curvature of the lowest and second-lowest bands of the $\phi_{\rm seed}=0$ solution, whose Chern numbers are $-2$ and $1$, respectively.}
    \label{fig:HF_band_berry_phi0_pi}
\end{figure}

\begin{figure}[htbp]
    \centering
    \includegraphics[width=1.0\linewidth]{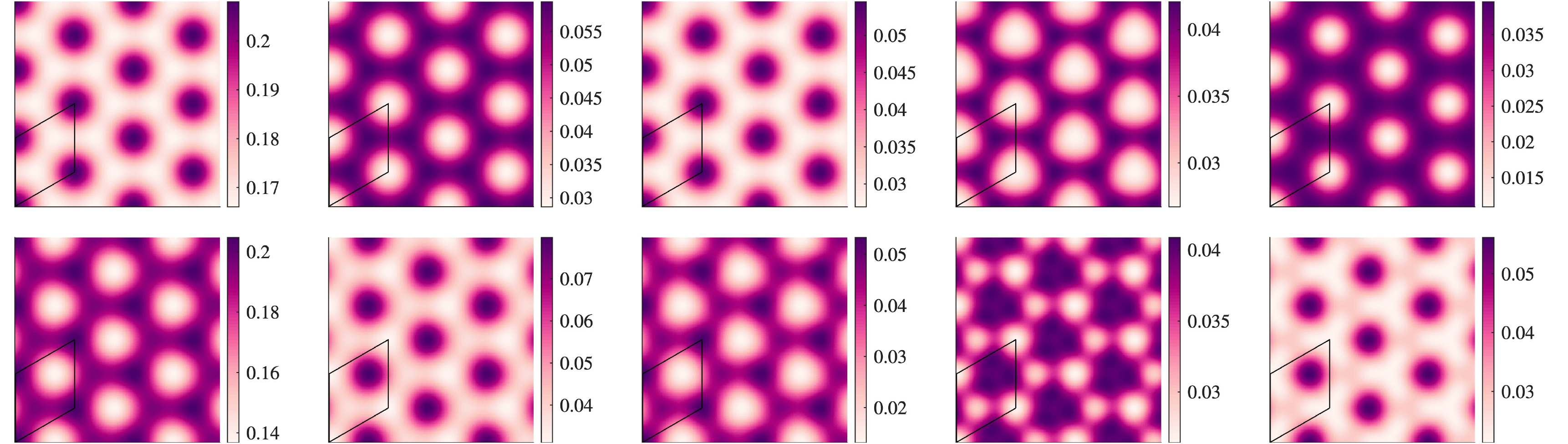}
    \caption{Layer-resolved real-space densities of the converged two-LL Hartree--Fock states at $\nu=1/3$, with $V_m=\SI{20}{meV}$ applied to the remote layer, physical moir\'e phase $\phi=0$, and dielectric constant $\epsilon=5$. The top and bottom rows are initialized with $\phi_{\rm seed}=0$ and $\phi_{\rm seed}=\pi/3$, respectively. Within each row, the five panels show layers $l=1,\ldots,5$ from left to right; the parallelogram outlines one moir\'e unit cell. Each panel has its own density color scale.}
    \label{fig:HF_density_phi0_pi_V20}
\end{figure}

As described in the main text, the same flux fraction can support distinct self-consistent HF solutions.
In our calculations, the solution reached by the iteration depends on the initial density distribution. We generate that initial density using a seed Hamiltonian with phase $\phi_{\rm seed}$, but the Hamiltonian used in every subsequent HF iteration always has the physical phase $\phi=0$.
Figures~\ref{fig:CF_band}(b), \ref{fig:HF_band_berry_phi0_pi}, and \ref{fig:HF_density_phi0_pi_V20} retain the two lowest conduction LLs. For Fig.~\ref{fig:CF_band}(b), we use $\phi_{\rm seed}=\pi/3$ and obtain a solution whose occupied band has Chern number $C=1$.
Figure~\ref{fig:HF_band_berry_phi0_pi} compares this solution with the branch obtained from $\phi_{\rm seed}=0$, whose occupied band has $C=-2$, and shows the Berry-curvature distributions of the two lowest bands of the latter branch.

Figure~\ref{fig:HF_density_phi0_pi_V20} shows the corresponding layer-resolved real-space density profiles for both solutions. These profiles show that the HF state preserves a moir\'e-scale modulation while redistributing charge among the graphene layers. 

\begin{figure}[htbp]
    \centering
    \includegraphics[width=0.48\linewidth]{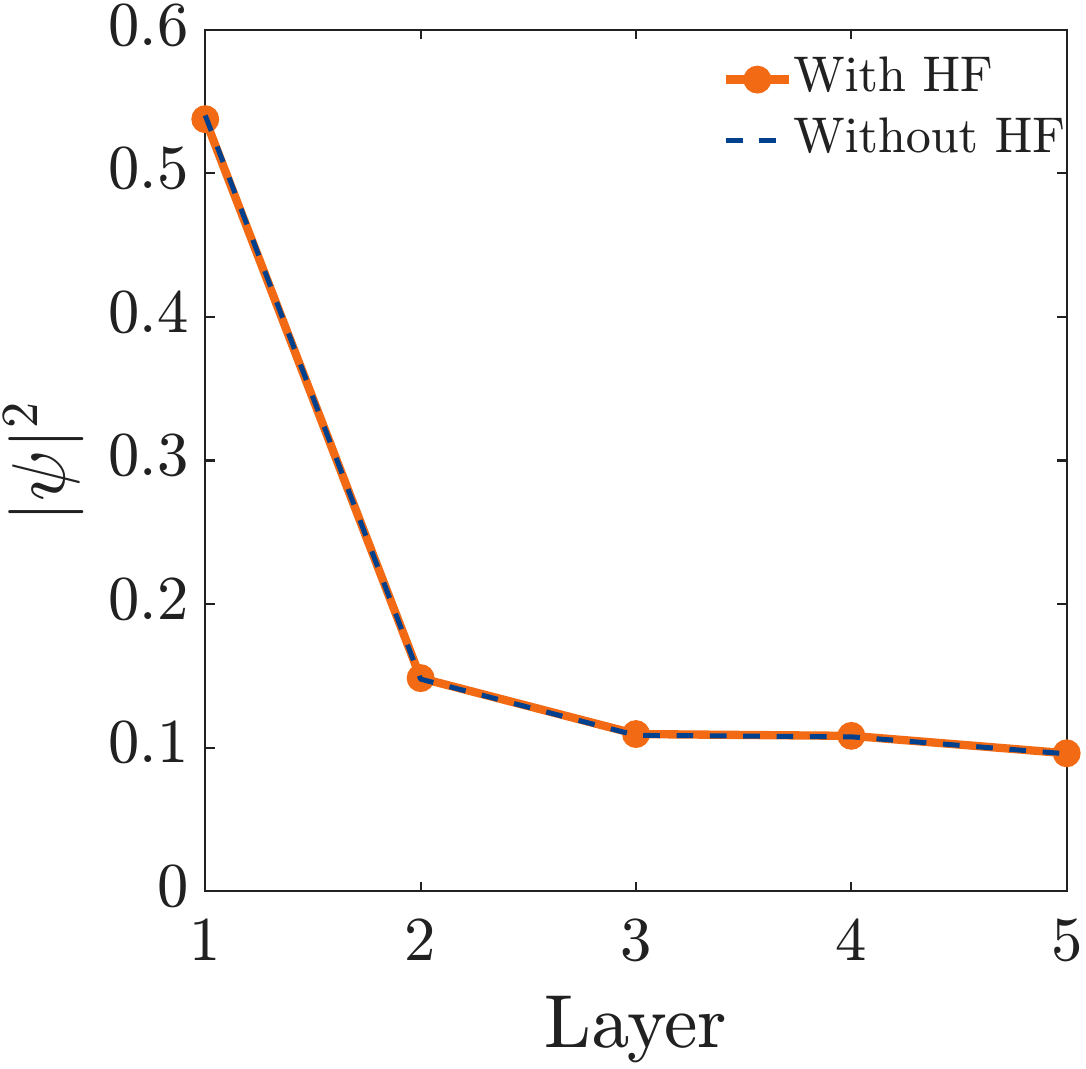}
    \caption{Layer distribution of the occupied CF magnetic Bloch band at $\nu=1/3$, with HF interactions (orange solid line with dots) and without HF interactions (dark blue dashed line). The plotted $|\psi|^2$ is the probability on each layer, summed over sublattices and orbital components and averaged over the magnetic Brillouin zone; each curve sums to one over the five layers. Both calculations use the same two-LL projected basis, momentum mesh, $u_D=\SI{25}{meV}$, $V_m=\SI{20}{meV}$ on layer $l_0=5$, and physical moir\'e phase $\phi=0$. The HF solution has $\epsilon=5$ and is initialized with $\phi_{\rm seed}=\pi/3$. The two curves nearly coincide.}
    \label{fig:HF_layer_distribution}
\end{figure}

The average layer distribution of the occupied band is compared before and after HF self-consistency in Fig.~\ref{fig:HF_layer_distribution}. For the $\phi_{\rm seed}=\pi/3$ solution used in Fig.~\ref{fig:CF_band}(b), HF interactions change the layer weights only weakly: the weight on layer $l=1$ decreases from $0.5404$ to $0.5377$, with the remaining weight redistributed among the other four layers. Thus, the interaction-driven band inversion largely preserves the layer distribution of the occupied CF band.

\begin{figure}[htbp]
    \centering
    \includegraphics[width=1.0\linewidth]{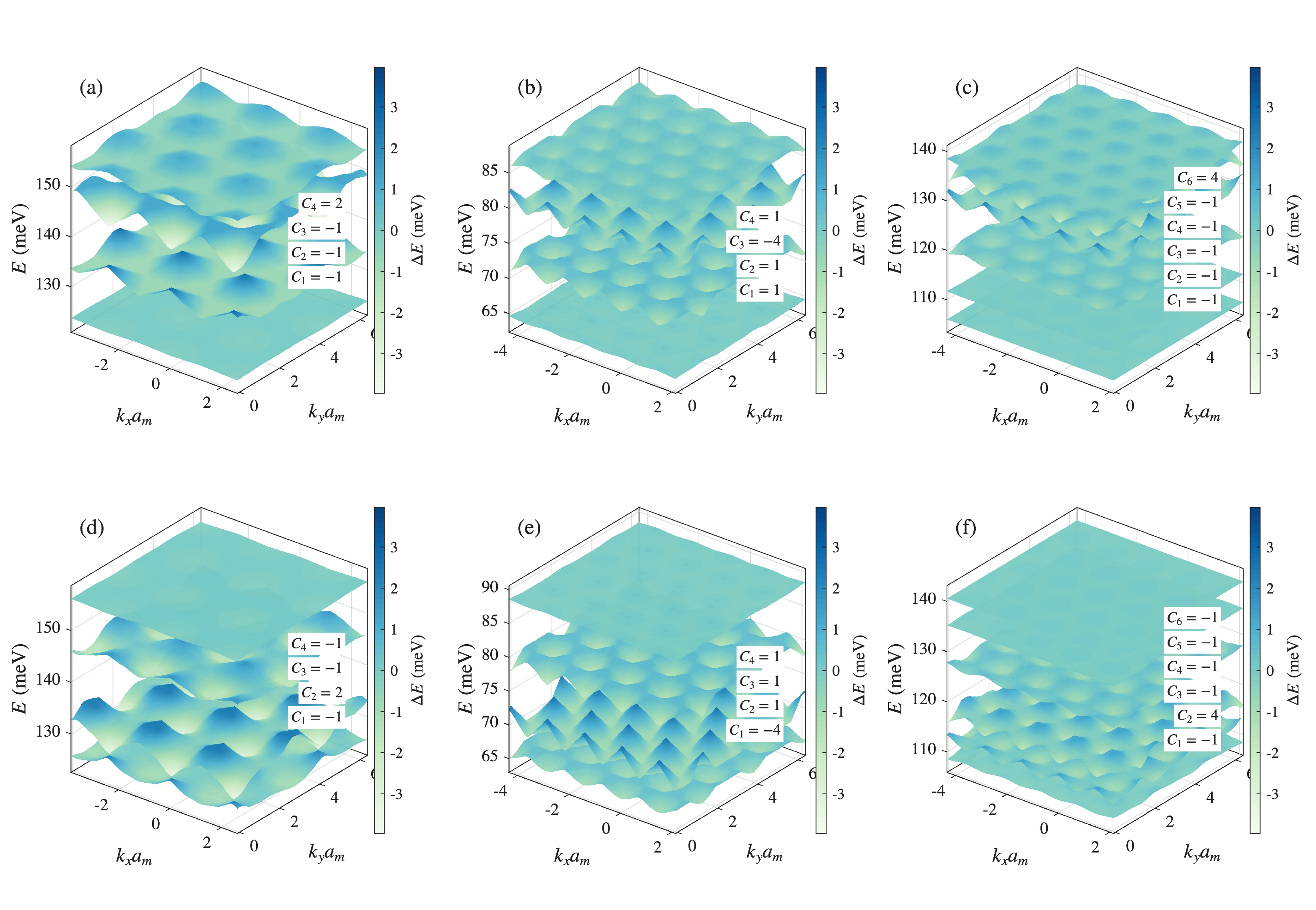}
    \caption{Single-LL CF magnetic Bloch bands for additional filling fractions, with physical moir\'e phase $\phi=0$, $V_m=\SI{20}{meV}$ on the remote layer, and dielectric constant $\epsilon=5$. From left to right, the columns correspond to $(p,q)=(4,3)$, $(4,5)$, and $(6,5)$, or $\nu=2/3$, $2/5$, and $3/5$, respectively. All panels show converged HF bands. Panels (a)--(c) are initialized with $\phi_{\rm seed}=0$, $\pi/3$, and $0$, while panels (d)--(f) are initialized with $\phi_{\rm seed}=\pi/3$, $0$, and $\pi/3$, respectively. Every band is labeled by its Chern number; at flux $p/q$, the lowest $p/2$ bands are occupied. The color scale shows $\Delta E_n({\bf k})\equiv E_n({\bf k})-\overline{E}_n$, the energy of each band relative to its mean.}
    \label{fig:HF_multiflux_end_bands_chern}
\end{figure}

We also calculate single-LL HF solutions for the filling fractions $\nu=2/5$, $3/5$, and $2/3$, shown in Fig.~\ref{fig:HF_multiflux_end_bands_chern}. The same mechanism of moir\'e-induced band reconstruction and gap opening remains operative. Both rows show converged branches initialized with complementary seed phases. Results for both seed phases, including their occupied-band Chern numbers and energies, are collected in Table~\ref{tab:HF_multiflux_energies}.

\clearpage
\begin{table*}[t]
    \caption{Dense-mesh Hartree--Fock results at several filling fractions, computed on $9\times(9q)$ momentum grids. The single-particle, Coulomb, total, and relative energies are quoted per particle. For each fraction, relative energy equal to zero identifies the lowest-energy state among the listed seed phases. The Hall conductivity is obtained from $\sigma_{xy}=-[C/(2C+1)]e^2/h$.}
    \label{tab:HF_multiflux_energies}
    \centering
    \begin{ruledtabular}
    \begin{tabular}{ccccrrrr}
    $\nu$ & $\phi_{\rm seed}$ & Occupied $C$ & $\sigma_{xy}/(e^2/h)$ & \multicolumn{4}{c}{$E/N$ (meV)} \\
    \cline{5-8}
    & & & & Single particle & Coulomb & Total & Relative energy \\
    \hline
    $1/3$ & $0$     & $-2$ & $-2/3$ &  74.651 & $-12.324$ &  62.327 & 0.118 \\
    $1/3$ & $\pi/3$ & $+1$ & $-1/3$ &  75.423 & $-13.214$ &  62.209 & 0.000 \\
    $2/5$ & $0$     & $-3$ & $-3/5$ &  90.764 & $-13.236$ &  77.528 & 0.000 \\
    $2/5$ & $\pi/3$ & $+2$ & $-2/5$ &  91.357 & $-13.715$ &  77.643 & 0.115 \\
    $3/5$ & $0$     & $-3$ & $-3/5$ & 141.259 & $-16.633$ & 124.627 & 0.000 \\
    $3/5$ & $\pi/3$ & $+2$ & $-2/5$ & 141.491 & $-16.647$ & 124.845 & 0.218 \\
    $2/3$ & $0$     & $-2$ & $-2/3$ & 157.075 & $-17.271$ & 139.804 & 0.000 \\
    $2/3$ & $\pi/3$ & $+1$ & $-1/3$ & 157.506 & $-17.597$ & 139.909 & 0.105
    \end{tabular}
    \end{ruledtabular}
\end{table*}
\clearpage

\section{Moir\'e-proximate regime with moir\'e on the top layer}

\begin{figure}[htbp]
    \centering
    \includegraphics[width=1.0\linewidth]{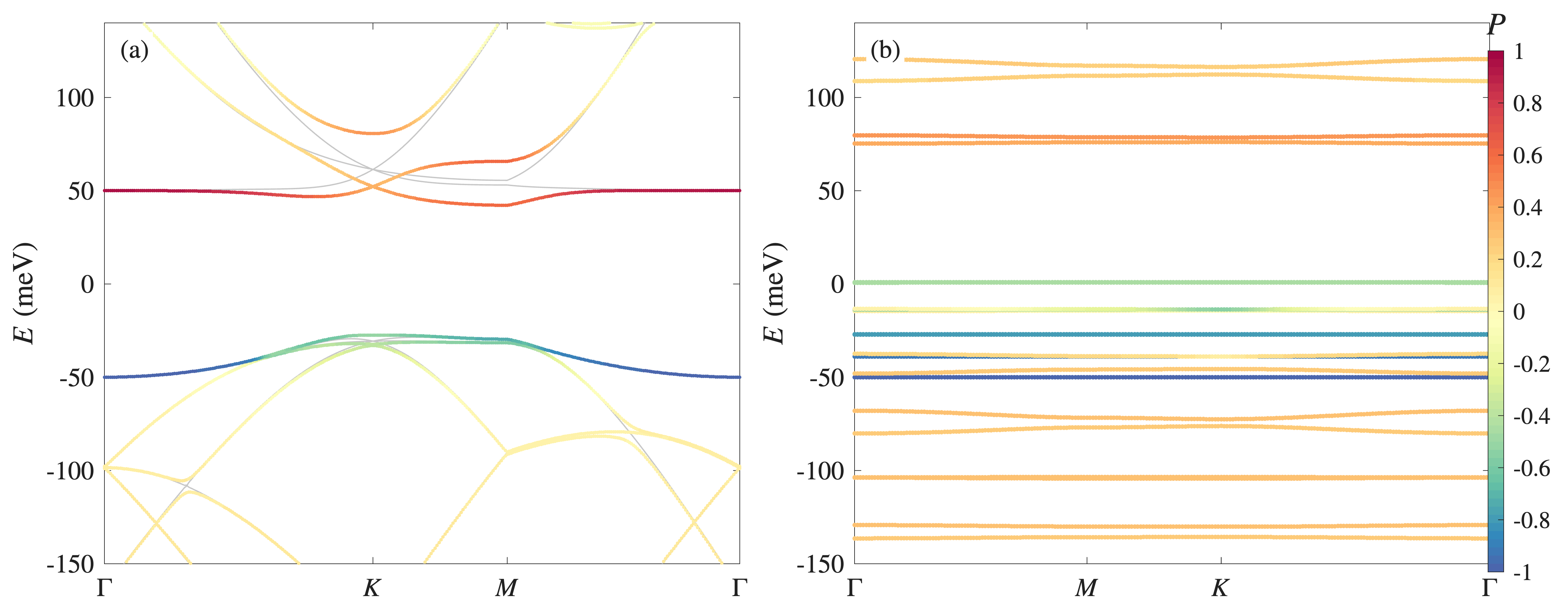}
    \caption{Top-layer moir\'e regime at $V_m=\SI{20}{meV}$ and $\phi=0$. Panel (a) shows the moir\'e-reconstructed electronic band structure with the $V_m=0$ reference in gray. Panel (b) shows the corresponding composite-fermion magnetic Bloch spectrum for the same top-layer moir\'e configuration.}
    \label{fig:supp-top-moire-proximate}
\end{figure}

Fig.~\ref{fig:supp-top-moire-proximate} compares the single-particle moir\'e band structure and the corresponding composite-fermion magnetic Bloch spectrum when the modulation is applied on the top layer ($l_0=1$). 
In the electronic problem, the moir\'e potential has a larger bare effect because the low-energy conduction states are already concentrated near that layer. 
After flux attachment, the same potential splits the lowest CF Landau level into magnetic subbands
separated by a visible gap, comparable to that produced by the remote-layer
potential in the main text.

The moir\'e-induced CF gap therefore does not rely on a particular microscopic limit. When the
moir\'e potential acts on the remote layer, its effect on the electronic bands is weak but is
amplified in the composite-fermion description through the layer-delocalized Landau-level
wave functions. When it acts on the top layer, the electronic bands are already strongly
modulated, and the CF spectrum is gapped as well.

In the moir\'e-proximate configuration, the moir\'e potential already modulates the electronic
bands strongly at the single-particle level. It may therefore favor electronically ordered
phases, such as topologically trivial charge density waves, which are not captured by the CF
mean-field theory. Such competing states may account for the insulating phases other than
FCIs observed when the moir\'e potential acts on the majority electron layer.

\end{document}